\documentclass[a4paper,11pt]{article}
\pdfoutput=1 

\usepackage{jcappub} 

\usepackage[T1]{fontenc} 
\usepackage{subfigure}
\usepackage{multirow}

\title{\texttt{SWIM}: Stochastic  Warm Inflation Module to generate and analyse Warm Inflationary power spectrum}

\author[a,1]{Umang Kumar\note{Corresponding author}}
\author[a]{and Suratna Das}

\affiliation[a]{Department of Physics, Ashoka University, Rajiv Gandhi Education City, Rai, Sonipat: 131029, Haryana, India}

\emailAdd{umang.kumar$\_{}$phd21@ashoka.edu.in}
\emailAdd{suratna.das@ashoka.edu.in}

\abstract{
 Numerical analysis to determine the form of the scalar power spectrum in Warm Inflationary paradigm is inevitable. One further needs numerical techniques to analyse any Warm Inflation model with the current observational data through the MCMC codes that are available publicly, like \texttt{COSMOMC} or \texttt{Cobaya}. We present \texttt{SWIM} (Stochastic Warm Inflation Module) written in \texttt{C++} and \texttt{Python}, that not only helps generate the Warm Inflationary scalar power spectrum, either semi-analytically or fully numerically, but also is integrated with \texttt{Cobaya} enabling the user to constrain the model parameters with current CMB data and thus to put any Warm Inflation model to test. \texttt{SWIM} numerically solves the standard stochastic perturbation equations of Warm Inflation without any approximations, uses machine learning techniques to speed up the MCMC analysis while analysing the fully numerical power spectrum that significantly reduces the computational cost, and is able to accommodate any Warm Inflation model with any form of inflationary potential and dissipative coefficient for numerical analysis. We show that \texttt{SWIM}, in most of the cases, outperforms other numerical codes on Warm Inflation that are designed to yield only the semi-analytical power spectrum as far as the runtimes are concerned. We further point out that there can be situations where the semi-analytical way of determining the scalar power spectrum in Warm Inflation can fall short, and one needs the full numerical power spectrum for parameter estimation given the observational data. In such cases, \texttt{SWIM} is the only code available so far that is designed to perform the task. Hence, \texttt{SWIM} offers a complete numerical platform for thorough analysis of Warm Inflation models against the current cosmological data. \texttt{SWIM} has been made publicly available at https://github.com/umg-kmr/SWIM. 
}

\begin{document}
\maketitle
\flushbottom

\section{Introduction}
\label{intro}

Warm Inflation (WI) \cite{Berera:1995ie} is a promising alternative to the standard inflationary paradigm (which we will refer to as Cold Inflation (CI) henceforth) where the couplings of the inflaton field with other degrees of freedom are not assumed to be suppressed, and through those couplings the inflaton field continuously dissipates its energy to a concurrent subdominant radiation bath during the inflationary period. Such dissipative dynamics was shown to be able to maintain a constant subdominant radiation bath during WI that helps the Warm Inflationary phase to gracefully exit into a radiation dominated epoch without invoking a need for a post-inflationary reheating phase, dynamics of which are still largely unknown. Evading the need of reheating post inflation is one of the advantages of WI dynamics over the standard CI scenarios. 

Owing to its very construction, WI has several other advantages. First of all, the dissipative mechanism results in extra friction in the inflaton's dynamics during WI apart from the Hubble friction that is already present due to the background expansion. When the friction due to dissipation takes over the Hubble friction, slow-roll dynamics can take place  in WI even when the potential becomes very steep. WI along very steep potentials have been studied in the literature in \cite{Das:2020xmh, Das:2025teu}. It is common knowledge that CI calls for flat enough potentials for slow-roll, and such flat potentials should also have a minimum where the field can oscillate during the reheating phase. Both these features of the CI inflaton potentials are not essential for WI to take place, and thus WI can accommodate a large variety of scalar potentials for realising inflation that are otherwise not suitable for CI dynamics. Secondly, WI in general takes place at lower energy scales than CI yielding lesser tensor amplitude and suppressed tensor-to-scalar ratio than CI. Therefore, potentials ruled out in CI for generating large tensor-to-scalar ratios, like quadratic \cite{Berera:2025vsu} and quartic \cite{Bartrum:2013fia} monomial potentials, can become viable potentials for WI. Thirdly, strongly dissipative WI models naturally turn out to be of small-field nature as the field excursion during WI generally remains sub-Planckian \cite{Bastero-Gil:2019gao, Kamali:2019xnt, Das:2020xmh}, and PLANCK data prefers small-field inflationary models over the large-field ones \cite{Planck:2018jri}. Due to this very feature as well as the fact that WI can take place along steep potentials, WI, by construction, is in tune with the two Swampland conjectures \cite{Das:2018hqy, Das:2018rpg, Motaharfar:2018zyb, Das:2019hto}, namely the distance \cite{Ooguri:2006in} and the de Sitter conjecture \cite{Ooguri:2018wrx, Garg:2018reu}, which is rather difficult to accommodate in single field CI models \cite{Kinney:2018nny}. Thus, WI is better suited for UV-complete theories like String Theory as an inflationary paradigm than CI if the Swampland Conjectures stand the test of time. In the past decade, sound particle physics constructions of WI have been achieved in the literature, such as the Warm Little Inflaton model \cite{Bastero-Gil:2016qru}, the Minimal Warm Inflation model \cite{Berghaus:2019whh}, and the Effective Field Theory model \cite{Bastero-Gil:2019gao}. These studies have put WI on stronger footings as viable inflationary models. A few attempts have also been made to realise WI in String Theory motivated theories \cite{Chakraborty:2025yms, Chakraborty:2026eep}. Moreover, some WI models can naturally yield Primordial Black Holes (PBHs) without invoking the need to depart from slow-roll \cite{Arya:2019wck, Bastero-Gil:2021fac, Arya:2023pod} as in CI \cite{Ozsoy:2023ryl} (going beyond slow-roll in WI faces several challenges as has been explored in \cite{Biswas:2023jcd, Biswas:2024oje}). This is an attractive feature of WI as currently PBHs are considered to be contending Dark Matter candidates \cite{Green:2024bam}. Last but not the least, while, as reported in the literature, CI models struggle to meet the current observational bounds on the scalar spectral index $n_s$ \cite{Kallosh:2025ijd}  set by the recent ACT observations \cite{AtacamaCosmologyTelescope:2025blo, AtacamaCosmologyTelescope:2025nti}, WI models have been shown to satisfy those bounds naturally \cite{Berera:2025vsu, Das:2025teu, Chakraborty:2025jof, Chakraborty:2025yms, Chakraborty:2026eep}. For recent reviews on WI, see \cite{Kamali:2023lzq, Berera:2023liv}. 

However, as the evolution of the inflaton field is coupled with the radiation bath in WI rendering the system as a multicomponent fluid, the dynamics as well as the perturbation calculation become quite involved in WI \cite{Graham:2009bf, Bastero-Gil:2011rva}. One can still analytically solve for the inflaton perturbation equation but only approximately while assuming a variety of approximations and simplifications. One major assumption that goes into such analytical estimation of the scalar power spectrum is to assume that the dissipative coefficient has no dependence on the temperature of the thermal bath \cite{Graham:2009bf, Ramos:2013nsa}. Such an assumption helps the inflaton perturbation equation decouple from the radiation perturbation equation making the former suitable for being solved analytically. However, almost all the realistic particle physics models of Warm Inflation, including the ones mentioned above \cite{Bastero-Gil:2016qru, Berghaus:2019whh, Bastero-Gil:2019gao}, yield temperature dependent dissipative coefficients indicating that the approximate analytical solution of the inflaton perturbation equation and the scalar power spectrum obtained from those solutions are inaccurate. To account for the inaccuracies incorporated in the analytical scalar power spectrum due to such approximations, one first needs to determine the power spectrum numerically without making any approximations, and then multiply the analytical power spectrum with a correcting factor $G(Q)$ (where $Q$ is the ratio of the friction due to dissipation and Hubble friction term present in the equation of motion of the inflaton field in WI), which is the ratio of the accurate numerical power spectrum and the approximated analytical power spectrum, to obtain a semi-analytical form of the Warm Inflationary scalar power spectrum \cite{Graham:2009bf, Kamali:2023lzq}. Thus it is evident that numerically determining the scalar power spectrum is inevitable in WI. Moreover, due to fluctuation dissipation theorem, the perturbation equations in WI turn out to be stochastic in nature \cite{Gleiser:1993ea, Bastero-Gil:2011rva}, and thus, while determining the power spectrum numerically one needs to average over several realisations of the numerical solutions \cite{Kamali:2023lzq}.  However, it was recently shown in \cite{Ballesteros:2022hjk, Ballesteros:2023dno} that the Warm Inflationary stochastic perturbations equations can be cast into deterministic Fokker-Planck equations  of probability distributions of the perturbed quantities, and the power spectra arising from both these approaches are the same. This prescription showed a way to solve deterministic equations in order to obtain the scalar power spectrum in WI. 

Furthermore, to have an overall understanding about the functionality of any WI model it is important to constrain the model parameters given the current Cosmic Microwave Background (CMB) data. It is then needed to feed the WI semi-analytical scalar power spectrum into the publicly available MCMC codes, namely \texttt{COSMOMC} \cite{Lewis:2002ah, Lewis:2013hha} or \texttt{Cobaya} \cite{Torrado:2020dgo} (through \texttt{CAMB} \cite{Lewis:1999bs, Howlett:2012mh}). However, the form of the semi-analytical power spectrum of WI is quite intricate than the one we usually get in CI, and cannot be not readily inserted in \texttt{CAMB} for further analysis. Previously a semi-analytical technique was prescribed in \cite{Bastero-Gil:2017wwl, Arya:2017zlb} through which certain simple WI models can be fed into \texttt{CAMB}. A generalized numerical methodology was recently developed in \cite{Kumar:2024hju} through which one can feed in the semi-analytical power spectrum of any Warm Inflation model, i.e., with any form of inflaton potential and dissipative coefficient, into \texttt{CAMB}, and perform parameter estimation through \texttt{COSMOMC} or \texttt{Cobaya}. We further point out in this paper, for the first time to the best of our knowledge, that the semi-analytical power spectrum approach to constrain model parameters for WI falls short in cases where the $G(Q)$ factor has strong dependencies on the model parameters. It was highlighted in the literature that this correcting factor depends mainly on the form of the dissipation coefficient, and thus on $Q$, and depends only mildly on the form of the potential. However, as observed in this paper, $G(Q)$ can significantly depend on other model parameters as well, such as the overall normalization of the potential $V_0$, the relativistic degrees of freedom $g_*$, or any other parameters specific to the particle physics construction of that particular WI model. In such cases, using the semi-analytical form of the power spectrum for parameter estimation will lead to systematic biases as $G(Q)$ is calculated given only one set of parameter values. In such cases, it becomes inevitable to feed in the numerical power spectrum directly to the MCMC codes while varying all the relevant model parameters. 

Therefore, it is evident from the above discussion that the need for numerical analysis to analyse any Warm Inflation model is two-fold: (i) first of all, the WI scalar power spectrum needs to be evaluated numerically even if one chooses to analyse the model with a semi-analytical form of the power spectrum, and (ii) secondly, to feed the WI power spectrum, either the semi-analytical one or the full numerical one, into MCMC codes to constrain the model parameters with the latest CMB data. Here, we present one such numerical module, \texttt{SWIM} (Stochastic Warm Inflation Module), that serves both these purposes. Written in \texttt{C++} and \texttt{Python}, \texttt{SWIM} has three submodules that serves three distinct purposes: (i) The $G(Q)$ submodule can be used to generate a numerical form of the $G(Q)$ function required to obtain the semi-analytical form of the WI scalar power spectrum, (ii) The semi-analytical power spectrum submodule that can be used to constrain model parameters with the help of \texttt{CAMB} and \texttt{Cobaya} using the semi-analytical form of the scalar power spectrum, and (iii) The full numerical power spectrum submodule that can be used to feed in the numerical power spectrum directly into \texttt{CAMB} and then \texttt{Cobaya} bypassing the process of determining the $G(Q)$ factor which is redundant in this case. Therefore, \texttt{SWIM} is one complete package to analyse any Warm Inflation model. It is important to point out that two numerical codes, namely \texttt{WarmSPy} \cite{Montefalcone:2023pvh} and \texttt{WI2easy} \cite{Rodrigues:2025neh}, already exist in literature that can be used to generate the $G(Q)$ factor, but cannot be used readily with the MCMC codes. We will comment on the advantages and disadvantages of these two previous codes and compare them with \texttt{SWIM} in due course. 

We have organised the rest of the paper as follows. Sec.~\ref{WI-review} is dedicated to describe concisely the background evolutions, stochastic evolutions of the primordial perturbations, and the derivation of the semi-analytical scalar power spectrum in Warm Inflation.  In Sec.~\ref{SWIM-two-modules}, we describe the first two submodules of \texttt{SWIM} which are designed to generate the $G(Q)$ factor and analyse the semi-analytical power spectrum to constrain model parameters with the help of \texttt{CAMB} and \texttt{Cobaya}, respectively. We also compare the features of \texttt{SWIM} with that of the previous two numerical codes on WI, namely \texttt{WarmSPy} and \texttt{WI2easy}, and highlights its novelty in this section. In Sec.~\ref{SWIM-module-3}, we first motivate why in certain WI models the semi-analytically obtained scalar power spectrum turns out to be insufficient to constrain model parameters through MCMC analysis. We then describe the third submodule of \texttt{SWIM} that analyses the full numerical power spectrum in combination with \texttt{CAMB} and \texttt{Cobaya}, and verify its functionality with a specific example. In Sec.~\ref{summary},  we discuss the main features of \texttt{SWIM} and summarize our findings with concluding remarks. This paper also contains four appendices. In Appendix.~\ref{appendix:installation}, we furnish step-by-step instructions to install \texttt{SWIM}. The following three appendices provide user guidelines to use the three separate submodules of \texttt{SWIM}.


\section{Theoretical background of Warm Inflation}
\label{WI-review}

In this section we will briefly discuss the background evolutions of the inflaton field and the radiation bath, evolutions of scalar and thermal perturbations, and estimating the scalar power spectrum generated in Warm Inflation in a semi-analytical way. As the dynamics of the inflaton field is coupled with the evolution of subdominant radiation bath present during WI through dissipation, the background evolution as well as the evolution of the perturbations are significantly intricate in WI than they are in standard CI. 


\subsection{Evolution of background in WI}

In WI, the inflaton field and the concurrent, subdominant yet non-negligible radiation bath evolve simultaneously through the coupled equations:
\begin{eqnarray}
&&\ddot\phi+3H\dot\phi+V,_\phi=-\Upsilon(\phi,T)\dot\phi, \label{phi-evo}\\
&&\dot\rho_R+4H\rho_R=\Upsilon(\phi,T)\dot\phi^2 \label{rad-evo},
\end{eqnarray}
where the overdot denotes derivative with respect to cosmic time, $V,_\phi\equiv\partial V/\partial\phi$, and $\Upsilon$ is the rate of dissipation. It is customary to assume that the radiation bath remains near thermal equilibrium during WI so that a temperature $T$ can be assigned, and the radiation energy density can be written as 
\begin{eqnarray}
\rho_R=C_R T^4, \quad\quad C_R\equiv\frac{\pi^2}{30}g_*(T),
\label{rad-eqn}
\end{eqnarray}
where $g_*(T)$ is the number of relativistic degrees of freedom of the radiation bath at $T$. Apart from these two evolution equations, the background in WI evolves according to the Friedmann equation as 
\begin{eqnarray}
3M_{\rm Pl}^2H^2=\frac12\dot\phi^2+V(\phi)+\rho_R.
\label{friedmann}
\end{eqnarray}
One can see from Eq.~(\ref{phi-evo}) that due to dissipation, an extra friction term ($\Upsilon\dot\phi$) appears in the dynamics of the inflaton field apart from the usual Hubble friction ($3H\dot\phi$) due to the background expansion. Defining a dimensionless quantity $Q$, which is a ratio of these two friction terms,
\begin{eqnarray}
Q\equiv \frac{\Upsilon}{3H},
\label{Q-def}
\end{eqnarray}
WI models are often analysed in two limiting cases: (i) strong dissipative regime $(Q\gg1)$, i.e., when the inflaton dynamics is governed by the dissipative friction term, and (ii) weak dissipative regime $(Q\ll 1)$, i.e., when the dissipative friction is subdominant. However, in all cases WI requires $T>H$, which distinguishes the dynamics of WI from that of CI. 

In the coupled set of evolution equations, given in Eq.~(\ref{phi-evo}) and Eq.~(\ref{rad-evo}), energy flows from the inflaton field to the radiation bath through the dissipation coefficient $\Upsilon$. The form of the dissipative coefficient depends on the microphysics of the specific WI model that one considers, and it naturally depends on the inflaton field $\phi$ and the temperature $T$ of the radiation bath. Several such examples of dissipative coefficients can found in the literature, for example: (i) in a two-stage decay process where the inflaton field indirectly decays to light fields mediated by heavy bosonic fields yields $\Upsilon\propto T^3/\phi^2$ \cite{Moss:2006gt, Bastero-Gil:2010dgy, Bastero-Gil:2012akf}, (ii) an axionic field coupled to non-Abelian gauge fields results in a friction term via sphaleron transition which has the form $\Upsilon\propto T^3$ (this specific WI model is known as the Minimal Warm Inflation) \cite{Berghaus:2019whh}, (iii)  Warm Little Inflaton model, where the Nambu-Goldstone boson inflaton field has Yukawa couplings with light fermions, yields a dissipative coefficient of the form $\Upsilon\propto T$ \cite{Bastero-Gil:2016qru}. This motivates one to parametrize the dissipative coefficient as \cite{Kamali:2023lzq}
\begin{eqnarray}
\Upsilon(\phi,T)=\tilde C_\Upsilon M^{1-p-c}T^p\phi^c\equiv C_\Upsilon T^{p}\phi^c,
\label{generic-Ups}
\end{eqnarray}
where $\tilde C_\Upsilon$ is a dimensionless quantity that depends on the WI model and $M$ is a mass scale that is also specified by the model (making $C_\Upsilon$ a dimension-full constant). Here $p$ and $c$ are integers which can take positive or negative values. Though there is no bound on $c$, but $p$ should be within $[-4,4]$ as the stability of Warm Inflationary models suggest \cite{Moss:2008yb, delCampo:2010by, Bastero-Gil:2012vuu}. However, there are WI models where the dissipative coefficient doesn't take such a simple form. For example, in an effective field theory construction of WI demonstrated in \cite{Bastero-Gil:2019gao}, the dissipative coefficient takes the form
\begin{eqnarray}
\Upsilon\simeq C_\Upsilon g^4\frac{\tilde M^2T^2}{m^3(T)}\left[1+\frac{1}{\sqrt{2\pi}}\left(\frac{m(T)}{T}\right)^{3/2}\right]e^{-m(T)/T},
\label{EFT-Ups}
\end{eqnarray}
where $m(T)$ is the thermal mass of the light scalar fields that couple to the inflaton field in this model. It turns out that $m(T)=m_0^2+\alpha^2 T^2$, where $m_0$ is the vacuum mass of these light scalars and $\alpha$ being the coupling constant. Thus, in the high temperature limit ($T\gg m_0$), the dissipative coefficient effectively varies with the inverse power of the temperature ($\Upsilon\propto T^{-1}$), and in the low temperature limit  ($T\ll m_0$) one gets $\Upsilon \sim T^2$. In this particular model the dissipative coefficient doesn't depend on the inflaton field $\phi$. In another example, which is a variant to the Minimal Warm Inflation model where the axionic inflaton field couples to the Standard Model gluons, the $\phi$ dependence of the dissipative coefficient comes through its dependance on the Hubble parameter $H$. The dissipative coefficient in this model takes the form \cite{Berghaus:2025dqi}
\begin{eqnarray}
\Upsilon=\frac{\kappa(\alpha_gN_c)^5T^3}{2f_a^2}\left(\frac{1}{1+\kappa(\alpha_gN_c)^5(2\tilde N_f/N_c) \frac{T}{H(\phi,T)}}\right),
\end{eqnarray}
where $\alpha_g$ is the gluon gauge coupling, $\kappa$ is a dimensionless constant that depends on the dimension of the gauge group ($N_c$), the representation of the fermions ($N_f$) and the gauge coupling $\alpha_g$, $f_a$ is the axion decay constant, and $\tilde N_f=5$ is the effective number of quarks contributing to the dissipative coefficient. 

In our code \texttt{SWIM}, we will evolve the background equations as well as the perturbation equations w.r.t. the $e$-foldings $N$ rather than in cosmic time $t$. As $dN=Hdt$, one can write the background equations, Eq.~(\ref{phi-evo}), Eq.~(\ref{rad-evo}) (now in terms of $T$) and Eq.~(\ref{friedmann}), as 
\begin{eqnarray}
&&\phi''+\left[3(1+Q)+\frac{H'}{H}\right]\phi'+\frac{V,_\phi}{H^2}=0,\nonumber\\
&& T'+T=\frac{3H^2Q}{4C_RT^3}\phi'^2,\nonumber\\
&&H^2=\frac{2\left[V(\phi)+C_RT^4\right]}{6M_{\rm Pl}^2-\phi'^2},
\label{bkg-eqs}
\end{eqnarray}
where prime denotes derivative w.r.t. $N$.

To evolve these background equations numerically one requires to set the initial conditions in terms of specifying the initial values of $\phi$, $\phi'$ and $T$. The initial conditions are easy to set using the slow-roll approximated background equations. The slow-roll approximated forms of the above background equations can be written as 
\begin{eqnarray}
&&3(1+Q)\phi'+\frac{V,_\phi}{H^2}\simeq 0, \nonumber\\
&&T^4\simeq \frac{3H^2Q}{4C_R}\phi'^2, \nonumber\\
&&H^2\simeq \frac{V}{3M_{\rm Pl}^2}.
\end{eqnarray}
It is then easy to see that for given initial values of $\phi$ and $Q$ ($\phi_{\rm ini}$ and $Q_{\rm ini}$, respectively), one can determine the initial values of $\phi'$ and $T$:
\begin{eqnarray}
\phi'_{\rm ini}&=&\frac{M_{\rm Pl}^2V,_\phi(\phi_{\rm ini})}{V(\phi_{\rm ini})(1+Q_{\rm ini})},\nonumber\\
T_{\rm ini}&=&\left(\frac{V(\phi_{\rm ini})Q_{\rm ini}}{4C_RM_{\rm Pl}^2}\right)^{1/4}.
\label{ini-cond}
\end{eqnarray}
Moreover, one can also determine $C_\Upsilon$ with these initial conditions as 
\begin{eqnarray}
C_\Upsilon=\frac{\sqrt{3V(\phi_{\rm ini})}Q_{\rm ini}}{\tilde\Upsilon(\phi_{\rm ini}, T_{\rm ini})},
\label{ini-cups}
\end{eqnarray}
where $\tilde\Upsilon\equiv \Upsilon/C_\Upsilon$. The above three equations are used in \texttt{SWIM} in order to set initial conditions when the background equations are evolved numerically. 


\subsection{Evolution of scalar perturbations in WI}

Here we will outline how the scalar perturbations evolve in WI following the recent review \cite{Kamali:2023lzq} on WI. The perturbed FLRW metric including only the scalar perturbations can be written as 
\begin{eqnarray}
ds^2=-(1+2\alpha)dt^2-2a\partial_i\beta\,dx^idt+a^2\left[\delta_{ij}(1+2\varphi)+2\partial_i\partial_j\gamma\right]dx^idx^j,
\end{eqnarray}
where $\varphi$ is the gauge-dependent curvature perturbation, $\alpha$ is the lapse function and $\chi\equiv a^2\dot\gamma+a\beta$ is the  scalar shear. Other background scalar quantities, such as the inflaton field $\phi$, the radiation energy density $\rho_R$, and the radiation pressure density $p_R$ should also be linearly perturbed over their background values as 
\begin{eqnarray}
\phi({\mathbf x},t)&=&\bar\phi(t)+\delta\phi({\mathbf x},t),\nonumber\\
\rho_R({\mathbf x},t)&=&\bar\rho_R(t)+\delta\rho_R({\mathbf x},t),\nonumber\\
p_R({\mathbf x},t)&=&\bar p_R(t)+\delta p_R({\mathbf x},t).
\end{eqnarray}
Moreover, there would be momentum perturbations when one perturbs the stress-energy tensor (the momentum term identically vanishes in the FLRW background, and has no background value). We denote the scalar 3-momentum perturbation for the radiation fluid as $\Psi_R$.

In an interacting picture, where two or more fluids interchange energy and momentum, the total energy-momentum tensor $T^{\mu\nu}=\sum_\sigma T^{\mu\nu}_{(\sigma)}$ is covariantly conserved ($\nabla_\mu T^{\mu\nu}=0$), whereas for the individual energy-momentum tensors of each of the interacting fluids, one gets $\nabla_\mu T^{\mu\nu}_{(\sigma)}=Q^{\nu}_{(\sigma)}$, where $Q^{\nu}_{(\sigma)}$ (not to be confused with the dimensionless quantity $Q$ defined in  Eq.~(\ref{Q-def})) denotes the energy-momentum transfer to the $\sigma-$fluid. Thus, conservation of the total energy momentum tensor demands that $\sum_\sigma Q^\nu_{(\sigma)}=0$. In WI setup, two fluids, the inflaton field and the radiation bath, are coupled through energy exchange in the background level. It is easy to read from Eq.~(\ref{rad-evo}) that for the radiation bath $Q_{(R)}=\Upsilon\dot\phi^2$ (and thus it implies that $Q_{(\phi)}=-\Upsilon\dot\phi^2$). The evolutions of the energy and momentum perturbations of the radiation bath can be obtained by perturbing the energy-momentum conservation equation of the radiation fluid, and the corresponding evolution equations read in momentum space as 
\begin{eqnarray}
\delta\dot\rho_R&=&-H\left(4-\frac{3Q\dot\phi^2}{4\rho_R}\frac{\Upsilon,_TT}{\Upsilon}\right)\delta\rho_R+\frac{k^2}{a^2}\Psi_R+6HQ\dot\phi\delta\dot\phi+\Upsilon,_\phi\dot\phi^2\delta\phi+\frac{4\rho_R}{3}\kappa \nonumber\\
&&-3H\left(Q\dot\phi^2+\frac{4\rho_R}{3}\right)\alpha, \label{pert-1}\\
\dot\Psi_R&=&-3H\Psi_R-3HQ\dot\phi\delta\phi-\frac13\delta\rho_R-\frac{4\rho_R}{3}\alpha. \label{pert-2}
\end{eqnarray}
Here $k$ is the comoving wavenumber of the perturbations, $\Upsilon,_T$ and $\Upsilon,_\phi$ are partial derivatives of the dissipative coefficients w.r.t. temperature $T$ and inflaton field $\phi$ respectively, and $\kappa=3(H\alpha-\dot\varphi)+(k^2/a^2)\chi$. In addition to these perturbation equations, the evolution equation of the inflaton field fluctuations $\delta\phi$ can be obtained by perturbing the inflaton field equation and by adding stochastic quantum ($\xi_q$) and thermal ($\xi_T$) white noise terms (following the fluctuation-dissipation theorem) which reads as 
\begin{eqnarray}
\delta\ddot\phi&=&-3H(1+Q)\delta\dot\phi-\left(\frac{k^2}{a^2}+V,_{\phi\phi}+\Upsilon,_\phi\dot\phi\right)\delta\phi-\frac{\Upsilon,_TT\dot\phi}{4\rho_R}\delta\rho_R+\dot\phi(\kappa+\dot\alpha)\nonumber\\
&&-[3H(1+Q)\dot\phi+2V,_{\phi}]\alpha+\sqrt{\frac{2\Upsilon T}{a^3}}\xi_T+\sqrt{\frac{H^2(9+12\pi Q)^{1/2}(1+2n)}{\pi a^3}}\xi_q,
\label{pert-3}
\end{eqnarray}
where we have used the inflaton background equation given in Eq.~(\ref{phi-evo}) to arrive at this equation. Here, the normalization of the stochastic term $\xi_q$ is chosen such that the analytical power spectrum can reproduce CI power spectrum in the correct limits, as has been done in \cite{Rodrigues:2025neh}. The normalization of $\xi_q$ also helps choose vanishing initial conditions for the perturbations equations \cite{Rodrigues:2025neh} during numerical simulations.\footnote{In \cite{Ballesteros:2023dno}, a different approach was incorporated to determine the normalization of $\xi_q$ in the deterministic matrix approach introduced there. First the homogeneous equations were solved by setting the noise term to zero that ensures the Bunch-Davies initial conditions, and in the second iteration the equations were again solved now by setting the initial conditions to zero and keeping the noise term as a source to the homogeneous equations. It was pointed out that this approach slightly differs from the one where the normalization is set from the beginning in the moderate dissipative regime $Q\sim 0.1$.} Moreover, $n$ denotes the statistical distribution of the inflaton field due to the presence of the radiation bath. If and when the inflaton field thermalizes with the radiation bath, the distribution becomes that of the Bose-Einstein form ($n=1/[\exp[H/T]-1]$). However, whether the inflaton field will thermalize or not depends on the underlying microphysics that determines the scattering rate of the inflaton with other fields, and the inflaton field thermalizes only when such scattering rates are higher than the background expansion rate. But, in most of WI models, such analysis is not done, and as it was noted in \cite{Bastero-Gil:2017wwl, Ballesteros:2023dno, Kumar:2024hju}, inflaton's thermalization affects the curvature power spectrum only in the weak dissipative models. Thus, in our code \texttt{SWIM} we keep the option open whether to include the thermal distribution of inflaton field or not.  The noise terms in the above equation are taken to be of Gaussian nature with zero mean, $\langle\xi_T\rangle=0$ and $\langle\xi_q\rangle=0$, and having the two-point correlation functions as 
\begin{eqnarray}
\langle\xi_T({\mathbf k},t)\xi_T({\mathbf k}',t')\rangle&=&\delta(t-t')(2\pi)^3\delta({\mathbf k}-{\mathbf k}'),\\
\langle\xi_q({\mathbf k},t)\xi_q({\mathbf k}',t')\rangle&=&\delta(t-t')(2\pi)^3\delta({\mathbf k}-{\mathbf k}').
\end{eqnarray}
It is to be noted that during WI, when $T>H$, the thermal noise $\xi_T$ always dominates over the quantum noise term $\xi_q$. Thus the thermal fluctuations of the inflaton field become dominant in WI sourcing the adiabatic curvature perturbations, unlike in CI where the primordial fluctuations are primarily quantum in nature. 

Moreover, the thermal noise term $\xi_T$ appearing in the inflaton fluctuation equation (Eq.~\ref{pert-3})) must also appear in the perturbation equation of the radiation energy density (with an opposite sign) given in Eq.~(\ref{pert-1}) in order to conserve the total energy-momentum tensor. However, it was noted in \cite{Bastero-Gil:2014jsa} that there is an ambiguity in including the thermal noise term in the radiation part, as the appearance of thermal noise term either in the energy flux or in the momentum flux can help preserve the total stress-energy tensor. It was further noted in \cite{Bastero-Gil:2014jsa} that inclusion of the thermal noise term in the radiation energy perturbation equation only affects the results in the weak dissipative regime, while the strong dissipative dynamics is insensitive to the presence of this noise term. Thus, in our code \texttt{SWIM}, we include the thermal noise term in 
Eq.~(\ref{pert-1}) as 
\begin{eqnarray}
\delta\dot\rho_R&=&-H\left(4-\frac{3Q\dot\phi^2}{4\rho_R}\frac{\Upsilon,_TT}{\Upsilon}\right)\delta\rho_R+\frac{k^2}{a^2}\Psi_R+6HQ\dot\phi\delta\dot\phi+\Upsilon,_\phi\dot\phi^2\delta\phi+\frac{4\rho_R}{3}\kappa \nonumber\\
&&-3H\left(Q\dot\phi^2+\frac{4\rho_R}{3}\right)\alpha-\sqrt{\frac{2\Upsilon T}{a^3}}\dot\phi\xi_T,
\label{pert-4}
\end{eqnarray}
while keeping the option open for the user whether or not to include it in the numerical evolutions of the perturbation equations. The same option is given in \texttt{WI2easy} \cite{Rodrigues:2025neh} as well.

All the above three perturbation equations are written in a gauge-ready form. In \texttt{SWIM} we chose to work in the Newtonian gauge by setting $\varphi=-\alpha$ and $\beta=0=\gamma$. This yields $\chi=0$ and $\kappa=3(H\alpha+\dot\alpha)$. Thus in this gauge, the only relevant scalar metric perturbation is $\alpha$, evolution of which can be determined by perturbing the Einstein field equations, and can be written as 
\begin{eqnarray}
\dot\alpha=-H\alpha+\frac{1}{2M_{\rm Pl}^2}(\dot\phi\delta\phi-\Psi_R).
\label{pert-5}
\end{eqnarray}
Therefore, Eq.~(\ref{pert-2}), Eq.~(\ref{pert-3}), Eq.~(\ref{pert-4}), and Eq.~(\ref{pert-5}) represent the complete set of scalar perturbation equations in WI in the Newtonian gauge. 

We will now express all the relevant scalar perturbation equations as derivatives w.r.t. $N$:
\begin{eqnarray}
\delta\phi''&=&-\left[3(1+Q)+\frac{H'}{H}\right]\delta\phi'-\left(\frac{k^2}{a^2H^2}+\frac{V,_{\phi\phi}}{H^2}+\frac{\Upsilon,_\phi\phi'}{H}\right)\delta\phi-\frac{\Upsilon,_T\phi'}{4 H C_RT^3}\delta\rho_R+4\phi'\alpha'\nonumber\\
&&-\left(\frac{\Upsilon}{H}\phi'+\frac{2V,_\phi}{H^2}\right)\alpha+\sqrt{\frac{2\Upsilon T}{a^3H^3}}\xi_T+\sqrt{\frac{(9+12\pi Q)^{1/2}(1+2n)}{\pi a^3H}}\xi_q,\\
\delta\rho_R'&=&-\left(4-\frac{\Upsilon,_TH\phi'^2}{4C_RT^3}\right)\delta\rho_R+\frac{k^2}{a^2H}\Psi_R+2\Upsilon H\phi'\delta\phi'+\Upsilon,_\phi H\phi'^2\delta\phi+4C_RT^4\alpha' \nonumber\\
&&-\Upsilon H\phi'^2\alpha -\sqrt{\frac{2\Upsilon T H}{a^3}}\phi'\xi_T,\\
\Psi_R'&=&-3\Psi_R-\Upsilon\phi'\delta\phi-\frac{1}{3H}\delta\rho_R-\frac{4C_RT^4}{3H}\alpha,\\
\alpha'&=&-\alpha+\frac{1}{2M_p^2}\left(\phi'\delta\phi-\frac{\Psi_R}{H}\right),
\end{eqnarray}
where the stochastic terms are now written in terms of $N$, $\xi_T({\mathbf x},N)$ and $\xi_q({\mathbf x},N)$, having the two-point correlation functions as
\begin{eqnarray}
\langle\xi_T({\mathbf k},N)\xi_T({\mathbf k}',N')\rangle&=&\delta(N-N')(2\pi)^3\delta({\mathbf k}-{\mathbf k}'),\\
\langle\xi_q({\mathbf k},N)\xi_q({\mathbf k}',N')\rangle&=&\delta(N-N')(2\pi)^3\delta({\mathbf k}-{\mathbf k}').
\end{eqnarray}
We evolve the above set of perturbation equations in \texttt{SWIM}.

\subsection{The semi-analytic form of the scalar power spectrum in WI}

From the above discussion, it is evident that the perturbations in WI are stochastic in nature, and thus the power spectrum of the comoving curvature perturbations, ${\mathcal R}$, in WI is defined as 
\begin{eqnarray}
\Delta_{\mathcal R}=\frac{k^3}{2\pi^2}\langle|{\mathcal R}|^2\rangle,
\label{stoc-avg-ps}
\end{eqnarray}
where $\langle\dots\rangle$ denotes ensemble average over several realizations of the noise terms satisfying the above perturbation equations. As, during WI, the cosmological fluid is comprised of more than one component,  the inflaton field along with the radiation bath, the comoving curvature perturbations receives contributions from both the components, and can be written as \cite{Bastero-Gil:2011rva}
\begin{eqnarray}
{\mathcal R}=\sum_{i=\phi, R}\frac{\bar\rho_i+\bar p_i}{\bar\rho+\bar p}{\mathcal R}_i,
\end{eqnarray}
where ${\mathcal R}_i$ denotes the contribution to the comoving curvature perturbations from individual fluid given as 
\begin{eqnarray}
{\mathcal R}_i=\alpha-\frac{H}{\bar\rho+\bar p}\Psi_i.
\end{eqnarray}
In WI, $\bar\rho_\phi+\bar p_\phi=\dot\phi^2$, $\bar\rho_R+\bar p_R=(4/3)\bar\rho_R$, and the inflaton momentum perturbation is $\Psi_\phi=-\dot\phi\delta\phi$. This yields the total comoving curvature perturbation ${\mathcal R}$ as 
\begin{eqnarray}
{\mathcal R}=\alpha-\frac{H}{\bar\rho+\bar p}\left(\Psi_R-\dot\phi\delta\phi\right).
\label{comoving-R}
\end{eqnarray}
With certain approximations and assumptions it can be shown that the comoving curvature perturbation is proportional to the inflaton fluctuations, and thus determining the inflaton power spectrum can help estimating for the comoving curvature power spectrum. First of all, it is considered in WI that a constant radiation bath ($\dot\rho_R\approx 0$) is maintained throughout, yielding $\bar\rho_R\approx (3/4)Q\dot\phi^2$ (from Eq.~(\ref{rad-evo})). Secondly, it can be shown that on superhorizon scales the radiation momentum perturbations become proportional to the inflaton momentum perturbations ($\Psi_R\sim-Q\dot\phi\delta\phi$). Furthermore, neglecting the metric fluctuations ($\alpha\approx 0$, which is not a bad assumption because one can see from Eq.~(\ref{pert-5}) that on superhorizon scales $\alpha$ is suppressed by the square of Planck mass compared to $\delta\phi$ and $\Psi_R$), one arrives at 
\begin{eqnarray}
{\mathcal R}\approx \frac{H}{\dot\phi}\delta\phi,
\end{eqnarray}
yielding 
\begin{eqnarray}
\Delta_{\mathcal R}\approx \left(\frac{H}{\dot\phi}\right)^2\Delta_{\delta\phi},
\end{eqnarray}
when evaluated on superhorizon scales (or at Hubble crossing). Hence, an analytical estimation of the inflaton fluctuation power spectrum can yield an analytic form of the comoving curvature power spectrum in WI (as is also the case in CI). However, it can be seen from Eq.~(\ref{pert-3}), that the inflaton fluctuations are coupled with radiation fluctuations whenever the dissipative coefficient has dependence on temperature $T$ (note that, $\alpha\approx 0$ has already been assumed). Thus, to analytically solve for the inflaton fluctuations alone, one must assume that $\Upsilon$ has no dependence on $T$. This incorporates a significant assumption in solving the inflaton fluctuation equation analytically. With all these assumptions and approximations, one can arrive at a analytic form of the comoving curvature perturbations given as 
\begin{eqnarray}
\left.\Delta_{\mathcal R}\right|_{\rm analytical}= \left(\frac{H^2}{2\pi\dot\phi}\right)^2\left(1+2n+\frac{2\sqrt3\pi Q}{\sqrt{3+4\pi Q}}\frac{T}{H}\right).
\end{eqnarray}
One must note that to arrive at this form of the power spectrum one needs to further suppress terms with slow-roll coefficients as a first-order approximation. 

It is thus evident that to arrive at the above analytical form of the Warm Inflationary scalar power spectrum, a number of significant assumptions and approximations are called for. To account for those departures from the actual scenario, one then needs to multiply the above equation with a correcting factor $G(Q)$ to finally write the form of the semi-analytic scalar power spectrum in WI as 
\begin{eqnarray}
\left.\Delta_{\mathcal R}\right|_{\rm semi-analytical} =\left(\frac{H^2}{2\pi\dot\phi}\right)^2\left(1+2n	+\frac{2\sqrt3\pi Q}{\sqrt{3+4\pi Q}}\frac{T}{H}\right)G(Q),
\label{semi-ana-ps}
\end{eqnarray}
where this correcting factor $G(Q)$ can only be determined numerically, and, in fact, accounts for the differences between the actual full numerical solution and the approximated analytical form of the scalar power spectrum:
\begin{eqnarray}
G(Q)\equiv \frac{\left.\Delta_{\mathcal R}\right|_{\rm numerical}}{\left.\Delta_{\mathcal R}\right|_{\rm analytical}}.
\label{GQ}
\end{eqnarray}
Here, the numerical power spectrum $\left.\Delta_{\mathcal R}\right|_{\rm numerical}$ is defined as in Eq.~(\ref{stoc-avg-ps}), and the comoving curvature perturbation ${\mathcal R}$ is as given in  Eq.~(\ref{comoving-R}).
Hence, unlike CI, it is impossible to determine the primordial scalar power spectrum in WI without determining it numerically. The pre-existing codes, \texttt{WarmSPy} \cite{Montefalcone:2023pvh} and \texttt{WI2easy} \cite{Rodrigues:2025neh}, were primarily designed to determine the $G(Q)$ factor that appears in the semi-analytic form of the scalar power spectrum. While \texttt{WarmSPy} is designed to handle very limited forms of dissipative coefficients and inflaton potentials, \texttt{WI2easy}, on the other hand, is generalised enough to handle any form of dissipative coefficients and inflationary potentials. \texttt{SWIM}, too, can determine the $G(Q)$ factor for any given WI model with any form of dissipative coefficient and inflaton potential. However, the added advantage of \texttt{SWIM} is that so far it is the only code available that stores the stochastic scalar power spectrum as an output of the numerical analysis that can then be directly fed into MCMC codes (like \texttt{COSMOMC} and \texttt{Cobaya}) for parameter estimations. This turns out be an essential requirement for WI models where $G(Q)$ itself depends on model parameters. We will elaborate on this issue in a later section. In the following section, we will illustrate how \texttt{SWIM} can be used to generate the $G(Q)$ of the semi-analytical power spectrum of WI.


\section{Determining the semi-analytical power spectrum using \texttt{SWIM}}
\label{SWIM-two-modules}

The semi-analytical form of the primordial scalar power spectrum, given in Eq.~(\ref{semi-ana-ps}), is essential to analyse any WI model against data. Once the primordial power spectrum of a WI model is written in the form given in Eq.~(\ref{semi-ana-ps}), it can then be fed into \texttt{COSMOMC} or \texttt{Coboaya} following a generalized methodology developed in \cite{Kumar:2024hju} to estimate the best-fit parameter values of that WI model. It is evident from the previous discussions that determining the $G(Q)$ factor that appears in the semi-analytical power spectrum, one needs to determine the power spectrum numerically by solving the primordial perturbation equations through numerical evolutions. As, in WI, these perturbation equations turn out to be stochastic in nature due to the presence of the noise terms, $\xi_T$ and $\xi_q$, the numerical evolutions of these equations will be stochastic as well. \texttt{WarmSPy} \cite{Montefalcone:2023pvh} was the first publicly available code that was designed to solve these stochastic perturbation equations. However, as one is essentially interested in stochastically averaged two-point correlations of the primordial perturbations to determine the primordial power spectrum (see Eq.~(\ref{stoc-avg-ps})), it was shown in \cite{Ballesteros:2022hjk, Ballesteros:2023dno} that these stochastic averages can be interpreted as probability distributions of these perturbations that follow the Fokker-Planck equation that is essentially a deterministic equation. Thus, according to \cite{Ballesteros:2022hjk, Ballesteros:2023dno}, instead of solving the stochastic perturbation equations, one can solve the deterministic Fokker-Planck equation of the probability distributions of the two-point correlations of the primordial perturbations to obtain the primordial power spectrum in WI. This alternative deterministic formalism based on Fokker-Planck method is claimed to be equivalent to the standard stochastic approach followed in WI to determine the primordial spectrum, as shown in \cite{Ballesteros:2022hjk, Ballesteros:2023dno}. \texttt{WI2easy} \cite{Rodrigues:2025neh} has implemented this deterministic approach to solve for the WI power spectrum numerically to determine $G(Q)$. We have, however, developed the code \texttt{SWIM} such that it stochastically evolves the standard Warm Inflationary perturbation equations as prescribed in the previous section.

As there are already existing codes in the literature, namely \texttt{WarmSPy} and \texttt{WI2easy}, that solve the WI perturbation equations numerically to generate the $G(Q)$ factor required for expressing the semi-analytic WI power spectrum, it is important to highlight the differences between these existing codes and \texttt{SWIM}:
\begin{itemize}
\item {\bf Differences between \texttt{WarmSPy} and \texttt{SWIM}:} Though both the codes solve the standard stochastic Warm Inflationary perturbation equations, \texttt{WarmSPy} is quite restrictive in the sense that it is designed to analyse only specific inflationary potentials (namely Monomial potentials, Hilltop-like potentials, Natural Inflation potential and $\beta$-exponential potential) while assuming $Q$ to be constant throughout, i.e. $\Upsilon\propto H$. Therefore, \texttt{WarmSPy} cannot incorporate any specific form of realistic dissipative coefficients that are in general represented as given in Eq.~(\ref{generic-Ups}). For the same reason, the background quantities evaluated numerically in \texttt{WarmSPy} are also inaccurate. On the other hand, in \texttt{SWIM}, any model of WI with any form of inflationary potential and dissipative coefficient can be given as input, and thus not only the background quantities, but also the perturbations are more accurately calculated numerally in \texttt{SWIM} than in \texttt{WarmSPy}. Secondly, \texttt{WarmSPy} doesn't include the noise term $\xi_T$ in the radiation perturbation equation, whereas in \texttt{SWIM} this option is kept open (i.e., the user may or may not choose to include it in the numerical simulations done by \texttt{SWIM}). Thirdly, \texttt{WarmSPy} generates an approximate analytical expression of $G(Q)$ which merely enhances the visual appeal of the semi-analytical form of the power spectrum. However, no single parametric analytic expression can accurately capture the features of $G(Q)$ that is generated from the numerical power spectrum. Thus, the analytical form of $G(Q)$ generated by \texttt{WarmSPy} is always an approximated one.\footnote{It was also highlighted in \cite{Ballesteros:2023dno} that approximated analytical forms of $G(Q)$ can incorporate significant errors in computing the power spectrum and thus the authors advocated for treating the power spectrum fully numerically.} In \texttt{SWIM}, one obtains a table of discrete values of $G(Q)$ as a function of $Q$ in a numerical form without any fitting analytical expression. 

\item {\bf Differences between \texttt{WI2easy} and \texttt{SWIM}:} The major difference between \texttt{WI2easy} and \texttt{SWIM} is that \texttt{WI2easy} solves the deterministic Fokker-Planck equation to obtain the numerical power spectrum in order to determine $G(Q)$, whereas \texttt{SWIM} directly solves the standard stochastic perturbation equations of Warm Inflation. As a result, the numerical $G(Q)$ obtained from \texttt{WI2easy} is noise-free, whereas owing to the stochastic nature of the numerical calculations and the finite number of realizations, the raw output of \texttt{SWIM} exhibits residual noise. To obtain a robust and continuous representation, the data are subsequently smoothed and interpolated in \texttt{SWIM}, yielding a numerically stable $G(Q)$. Both \texttt{WI2easy} and \texttt{SWIM} do not yield any approximated analytical form of the $G(Q)$ function. 
\end{itemize}
We highlight the differences between all these three codes in Table~\ref{tab:comparison_table}. The last line of the table emphasises that \texttt{SWIM} is so far the only publicly available code that can be directly integrated with the MCMC codes, like \texttt{COSMOMC} or \texttt{Cobaya}. We will discuss this point elaborately in the next section. 
\begin{table}[h]
\centering
    \resizebox{\textwidth}{!}{ 
    \begin{tabular}{|c|c|c|c|}
        \hline
         & \texttt{WarmSPy} & \texttt{WI2easy} & \texttt{SWIM}  \\ \hline
        \textbf{Language}   & Python   & Mathematica   & C++ \& Python     \\ \hline
        \textbf{Formalism}   & Stochastic   & Fokker-Planck   & Stochastic    \\ \hline
               \textbf{$G(Q)$ Calculation}& Provides Analytical Fit  & Provides as a data file  & Provides as a data file    \\ \hline
               \textbf{Inclusion of $\xi_T$ in } & No  & Optional  & Optional    \\ 
               \textbf{ radiation perturbation equation} &   &   &     \\ \hline
         \textbf{Numerical power spectrum } & No  & Only semi-analytic  & Both semi-analytic    \\ 
         \textbf{ as an output} &  &  & as well as fully numerical    \\ \hline
                 \textbf{Bayesian Analysis Integration}  & No  & No  & Yes    \\ \hline
    \end{tabular}
    }
    \caption{Comparison between \texttt{WarmSPy}, \texttt{WI2easy} and \texttt{SWIM}}
    \label{tab:comparison_table}
\end{table}

While discussing the differences between \texttt{SWIM} and the other two existing codes, \texttt{WarmSpy} and \texttt{WI2easy}, it is important to note that \texttt{SWIM} is the only code where one can bypass determining $G(Q)$ and can obtain as an output the fully numerical power spectrum. Determination of $G(Q)$ for the semi-analytical power spectrum becomes futile when $G(Q)$ itself starts depending on the model parameters. The only way one can then deal with the parameter estimation of such a model is by feeding the fully numerical power spectrum into the MCMC codes. The next section will be devoted in explaining such cases as well as how to employ \texttt{SWIM} to determine the fully numerical power spectrum. Here, we will discuss the module of \texttt{SWIM} through which one can determine $G(Q)$, and compare it with outputs from \texttt{WI2easy}. As \texttt{WarmSPy} is very restrictive as well as not very accurate, we will not compare \texttt{SWIM} outputs with those of \texttt{WarmSPy} (also mainly because most of the models studied here cannot be analyzed by \texttt{WarmSPy} due to its very restrictive construction). 

\begin{figure}
\centering

\includegraphics[width=1.0\textwidth]{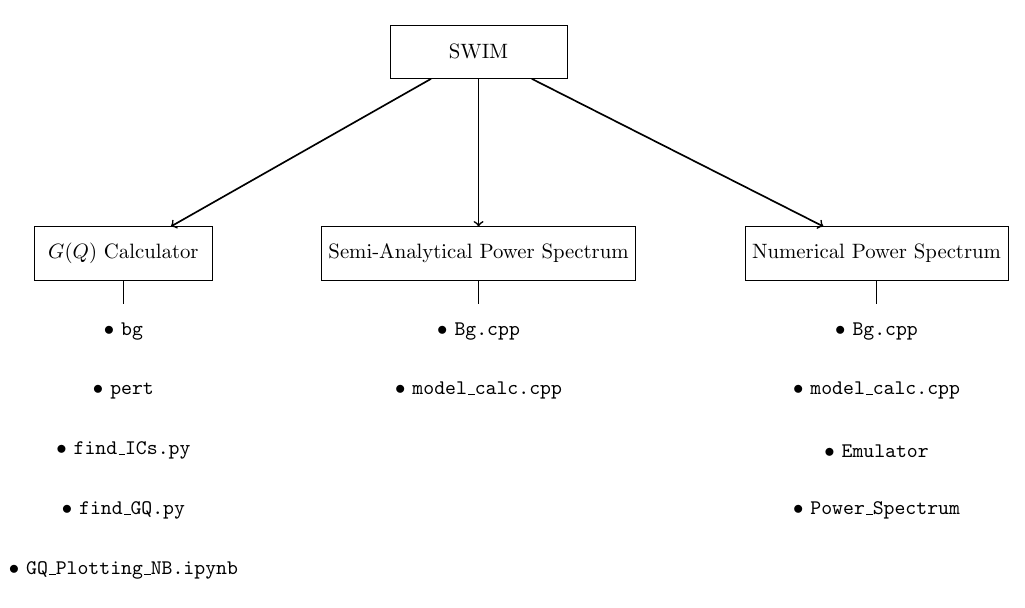}

\caption{Illustration of the submodules of \texttt{SWIM}.}
\label{flowchart:SWIM_Overview}

\end{figure}

\texttt{SWIM} has three main submodules to determine (i) the $G(Q)$ factor, (ii) the semi-analytical power spectrum, and (iii) fully numerical power spectrum as shown in Fig.~\ref{flowchart:SWIM_Overview}. In this section we will discuss the first two submodules that one can use to generate the semi-analytical power spectrum of WI. However, one can bypass both these modules and can generate the fully numerical power spectrum using the third submodule. We will discuss the third submodule in the next section. The instructions to install \texttt{SWIM} and the required environments are furnished in Appendix~\ref{appendix:installation}.


\subsection{The $G(Q)$ calculating submodule of \texttt{SWIM}}
\label{GQ-subsection}

This  submodule is organized into two subdirectories, \texttt{bg} and \texttt{pert}, along with two Python scripts, \texttt{find\_ICs.py} and \texttt{find\_GQ.py}, and a plotting notebook, \texttt{GQ\_Plotting\_NB.ipynb}. The user begins by specifying the inflationary model in the \texttt{model\_calc.cpp} files within both subdirectories. In the \texttt{bg} module, the user must define the WI model by specifying $V(\phi)$, $V_{,\phi}(\phi)$, and the dissipative coefficient $\Upsilon(\phi,T)$. In the \texttt{pert} module, which handles perturbation dynamics, the user must additionally provide $V_{,\phi\phi}(\phi)$ as well as the derivatives of the dissipative coefficient, $\Upsilon_{,T}(\phi,T)$ and $\Upsilon_{,\phi}(\phi,T)$. It is to note that the format in which the inflaton potential and its derivatives are to be inserted in \texttt{SWIM} is fixed: 
\begin{eqnarray}
V(\phi)=V_0f(\phi/M_{\rm Pl}), \label{pot-form}
\end{eqnarray}
where $V_0$ is the overall constant normalization of the potential that has a mass-dimension 4, and $f(\phi/M_{\rm Pl})$ is the dimension-less field-dependent part of the potential ($M_{\rm Pl}$ is set to be 1 throughout in \texttt{SWIM}). 
For dissipative coefficients that can be expressed in the generic form given in Eq.~(\ref{generic-Ups}), their derivatives are automatically computed internally as
\begin{eqnarray}
    \Upsilon_{,T}= p\, C_{\Upsilon} \phi^c T^{p-1}, \quad\quad\quad
    \quad \Upsilon_{,\phi} = c\, C_{\Upsilon} T^p \phi^{c-1},
\end{eqnarray}
and the user doesn't need to explicitly specify $\Upsilon_{,T}$ and $\Upsilon_{,\phi}$. For more general functional forms of $\Upsilon(\phi,T)$, the corresponding partial derivatives must be provided explicitly.
After specifying the model, the source code needs to be compiled using the \texttt{compile\_SWIM.sh} script, which generates shared libraries (\texttt{.so} files) in the \texttt{bg} and \texttt{pert} directories for use by the Python interface that includes \texttt{find\_ICs.py} and \texttt{find\_GQ.py}. 

The initial conditions $(\phi_i, Q_i)$ are obtained using the \texttt{find\_ICs.py} script, which determines values that yield a specified duration of inflation, controlled by the parameter \texttt{dur\_N} (default value: 60). The script computes these quantities over a user-defined range of $Q \in [Q_{\min}, Q_{\max}]$, using a parallelized root-finding algorithm. It uses the initial condition equations given in Eqs.~(\ref{ini-cond}) and Eq.~(\ref{ini-cups}) in order to evolve the background equations given in Eqs.~(\ref{bkg-eqs}). The script yields an output file named \texttt{ics.dat}, containing the initial field values $\phi_i$ corresponding to each $Q_i$. This step utilizes the \texttt{bg} submodule. The subsequent computation of $G(Q)$ is performed using the \texttt{find\_GQ.py} script which loads the \texttt{ics.dat} file generated by \texttt{find\_ICs.py}, and uses these initial conditions to compute the numerical power spectrum and evaluate the correction function $G(Q)$ using Eq.~(\ref{GQ}). The resulting output is saved in the file \texttt{GQ.dat}. Finally, the plotting notebook can be used to visualize the raw output of the $G(Q)$ calculator, as well as to smooth the results and save them for use in the semi-analytical power-spectrum module of \texttt{SWIM}. 

\begin{figure}
\centering

\includegraphics[width=1.0\textwidth]{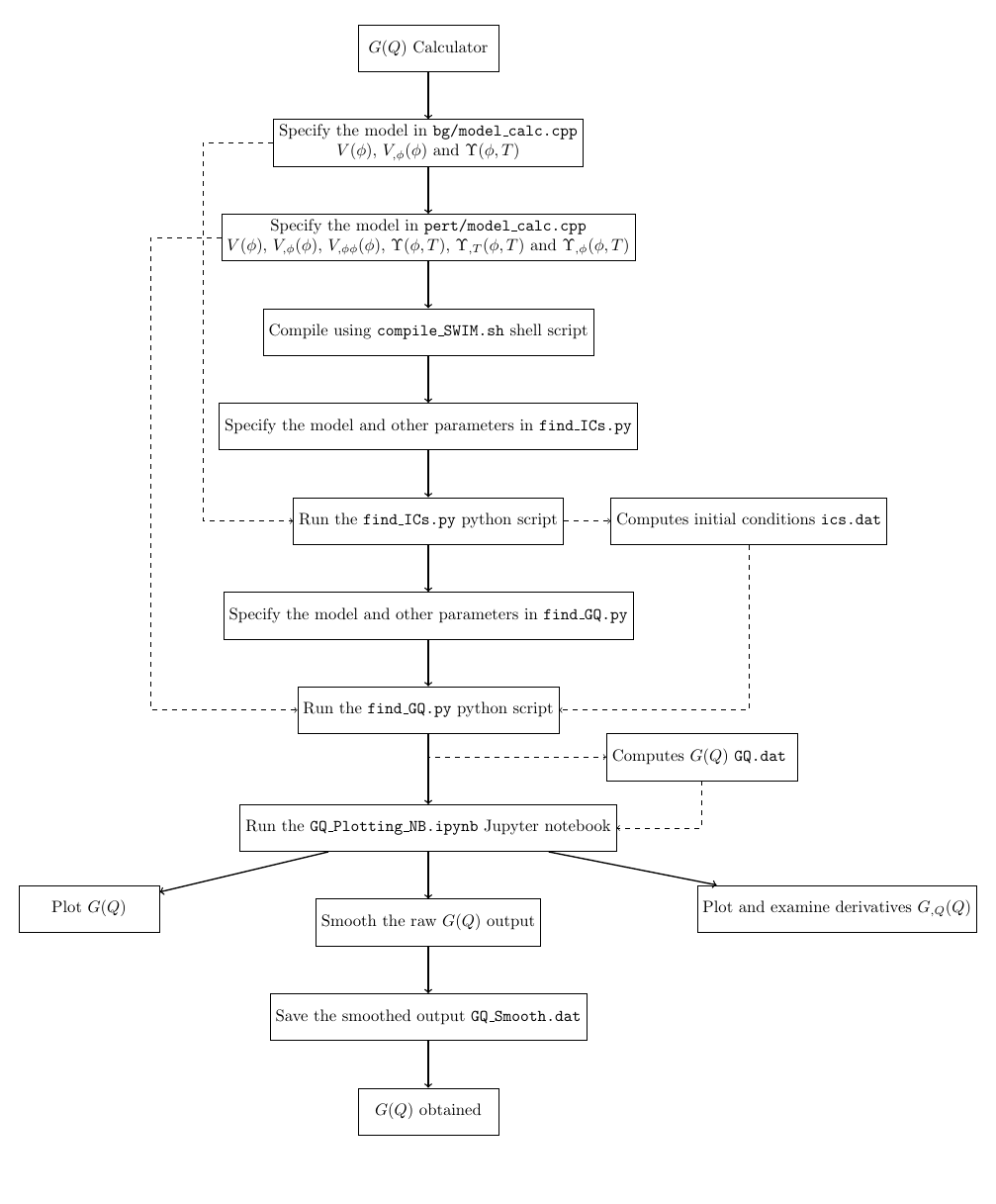}

\caption{Work flow of the  $G(Q)$ submodule }
\label{flowchart:GQ-Calc_PS}

\end{figure}

 A flowchart illustrating the workflow of the $G(Q)$ submodule of \texttt{SWIM} is shown in Fig.~\ref{flowchart:GQ-Calc_PS}. Here the solid path lines indicate the path to be followed by the user, while the dashed path lines indicate the internal simulations done by \texttt{SWIM}. We have  explained the inputs required from the user to run this submodule in Appendix~\ref{appendix:GQ}.


As mentioned above, we will compare the outputs of \texttt{SWIM} with that of \texttt{WI2easy}. In \cite{Rodrigues:2025neh} a number of potentials were considered as viable Warm inflationary potentials to study the performance of \texttt{WI2easy}, such as 
\begin{enumerate}
\item The quartic monomial potential \cite{Bartrum:2013fia}: $V(\phi)=\frac{V_0}{4}\left(\frac{\phi}{M_{\rm Pl}}\right)^4$,
\item The fibre type I inflation potential \cite{Chakraborty:2025yms}: \\ $V(\phi)=V_0\left[F-4e^{-\frac{\phi}{\sqrt3 M_{\rm Pl}}}+e^{-\frac{4\phi}{ \sqrt3 M_{\rm Pl}}}+R\left(e^{-\frac{2\phi}{\sqrt3  M_{\rm Pl}}}-1\right)\right]$, where $F$ and $R$ are dimensionless constants,
\item The cosine axion-type potential \cite{Montefalcone:2022jfw, Bastero-Gil:2026ypn}: $V(\phi)=V_0\left[1+\cos\left(\frac{\phi}{f_a}\right)\right]$, where $f_a$ is the axion decay constant,
\item A quadratic hilltop potential \cite{Rodrigues:2025neh}: $V(\phi)=V_0\left[1-\gamma\left(\frac{\phi}{M_{\rm Pl}}\right)^2\right]$, where $\gamma$ is a constant,
\item The hybrid inflation potential \cite{Berghaus:2019whh}: $V(\phi,\sigma)=\frac{1}{4\lambda}(M^2-\lambda\sigma^2)^2+\frac12m^2\phi^2+\frac12g^2\phi^2\sigma^2$, where $\sigma$ is the waterfall field. However, before the $\sigma$ field becomes dynamic, the potential essentially is of a single field $\phi$ as $\sigma$ sits at its true minimum $\sigma=0$. Thus, for $\phi>\phi_c$ ($\phi_c=M/g$ being the critical value of $\phi$ when $\sigma$ becomes dynamic) the effective potential is $V(\phi)=\frac{M^4}{4\lambda}+\frac{V_0}{2}\frac{\phi^2}{M_{\rm Pl}^2}$.
\end{enumerate}
These potentials were studied in combination with three most common Warm Inflationary dissipative coefficients: $\Upsilon=C_\Upsilon T^3/\phi^2$, $\Upsilon=C_\Upsilon T$, and $\Upsilon=C_\Upsilon T^3$. However, as we claimed that \texttt{SWIM} (as well as \texttt{WI2easy}) is capable of analyzing any form of dissipative coefficients that are not restricted to the generic form given in Eq.~(\ref{generic-Ups}), we will add to this list the EFT dissipative coefficient given in Eq.~(\ref{EFT-Ups}) in combination with a quadratic monomial potential of the form $V(\phi)=(V_0/2)(\phi/M_{\rm Pl})^2$ to study the performance of \texttt{SWIM} while comparing it against \texttt{WI2easy}. 

\begin{center}
\begin{figure}[!htb]
\subfigure[ Quartic monomial, $\Upsilon\propto T^3/\phi^2$ ]{\includegraphics[width=5cm]{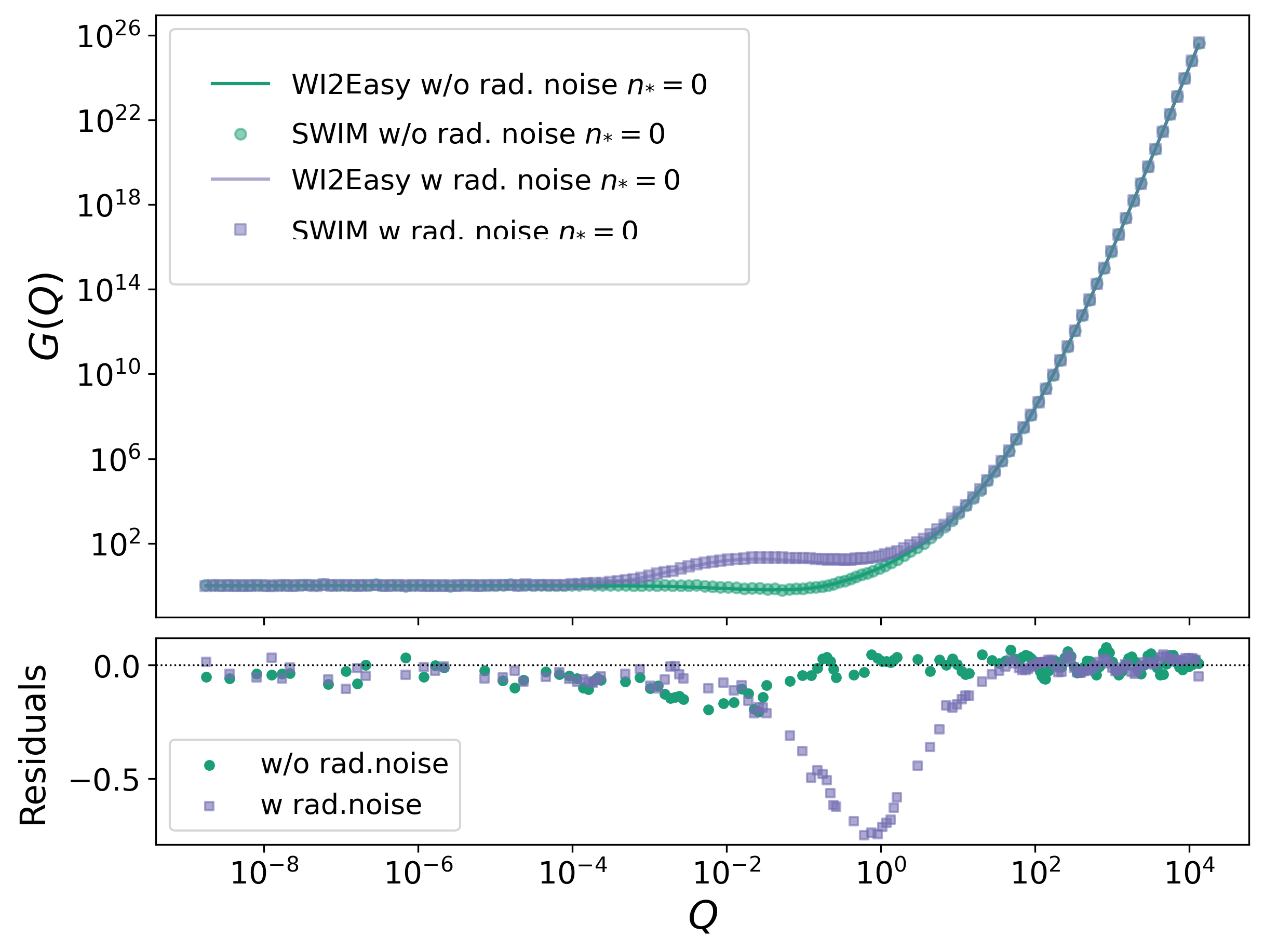}}
\subfigure[Quartic monomial, $\Upsilon\propto T$]{\includegraphics[width=5cm]{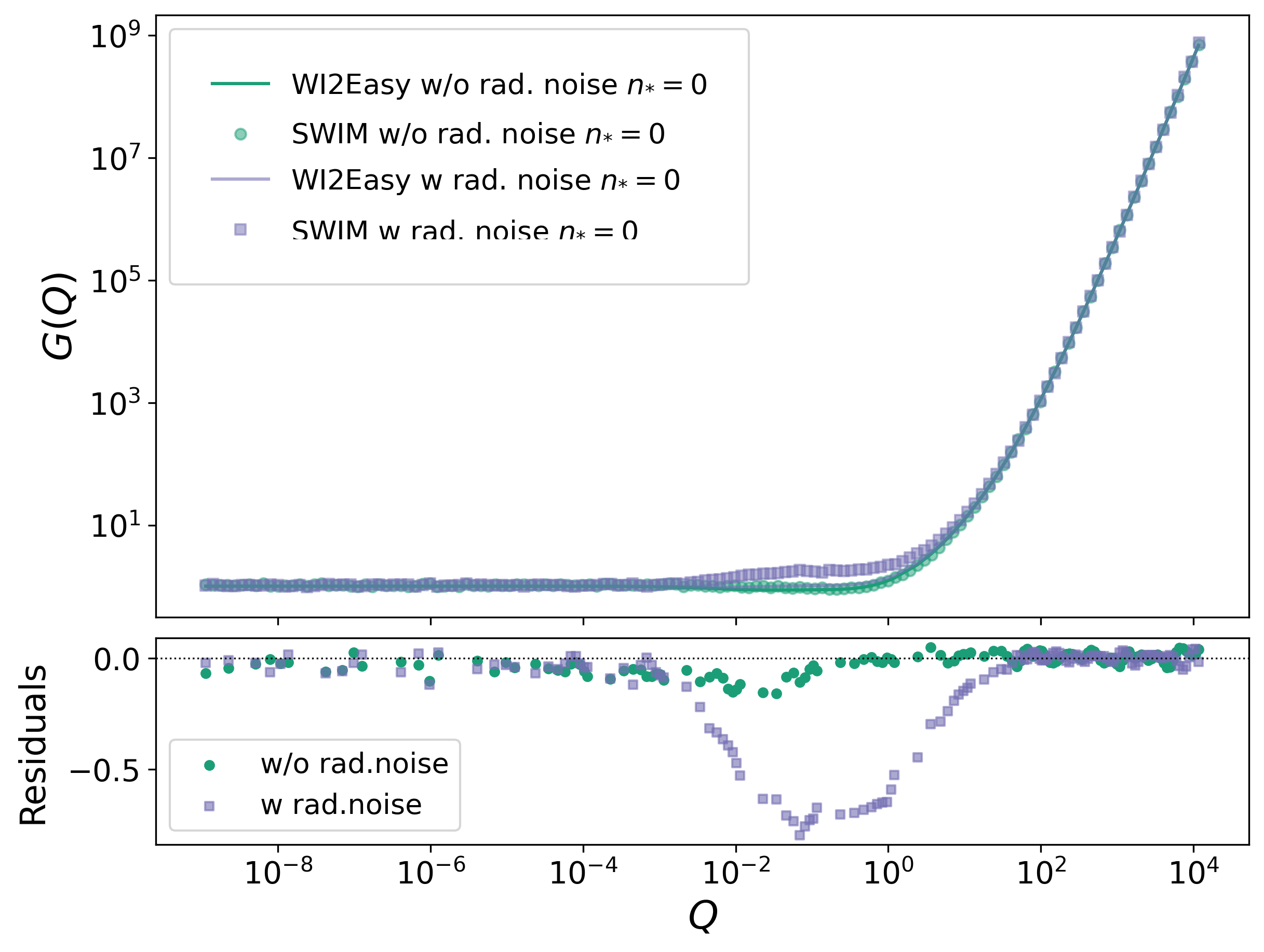}}
\subfigure[ Quartic monomial, $\Upsilon\propto T^3$]{\includegraphics[width=5cm]{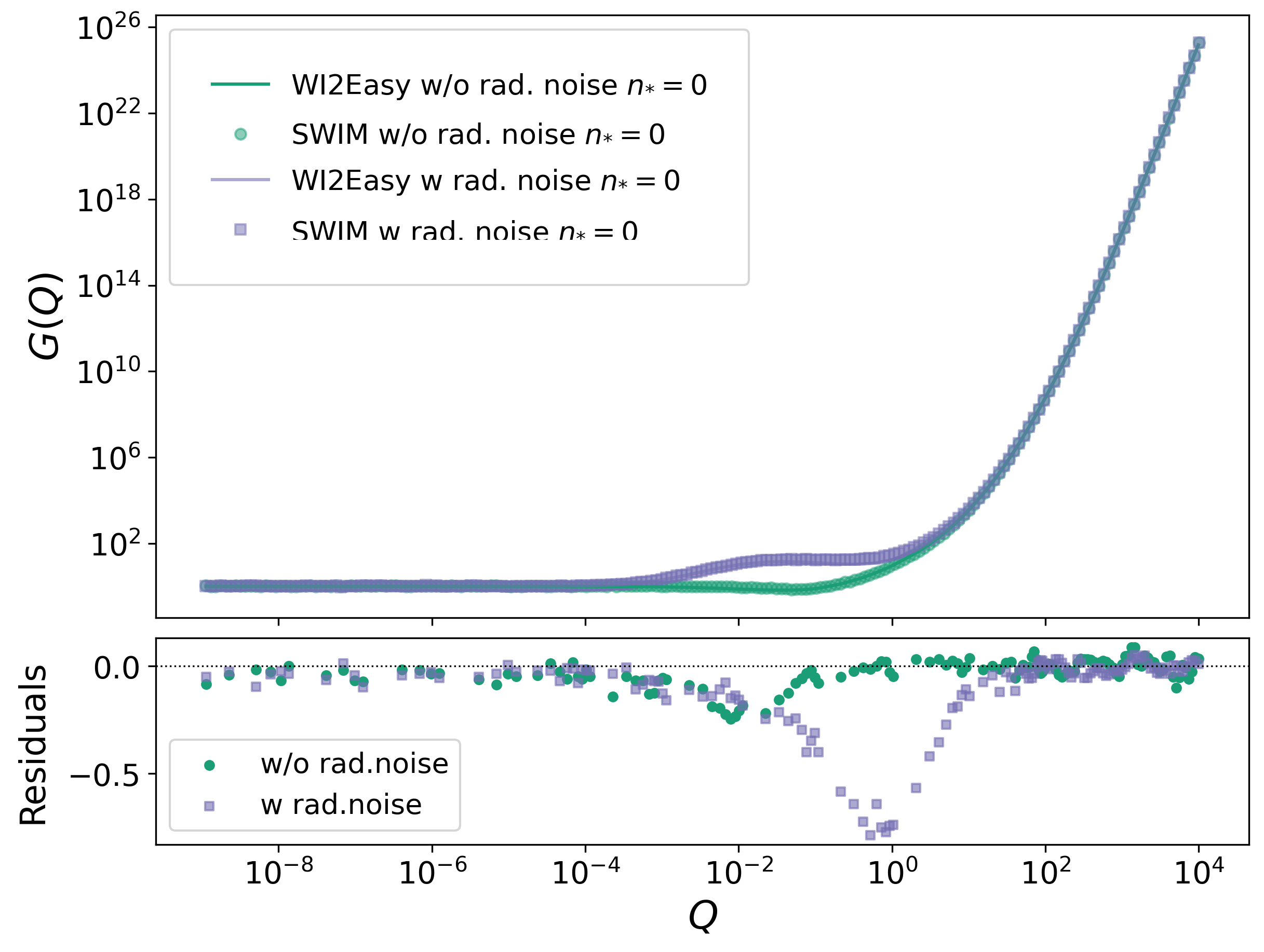}}\\
\subfigure[Fibre type I,   $\Upsilon\propto T$]{\includegraphics[width=5cm]{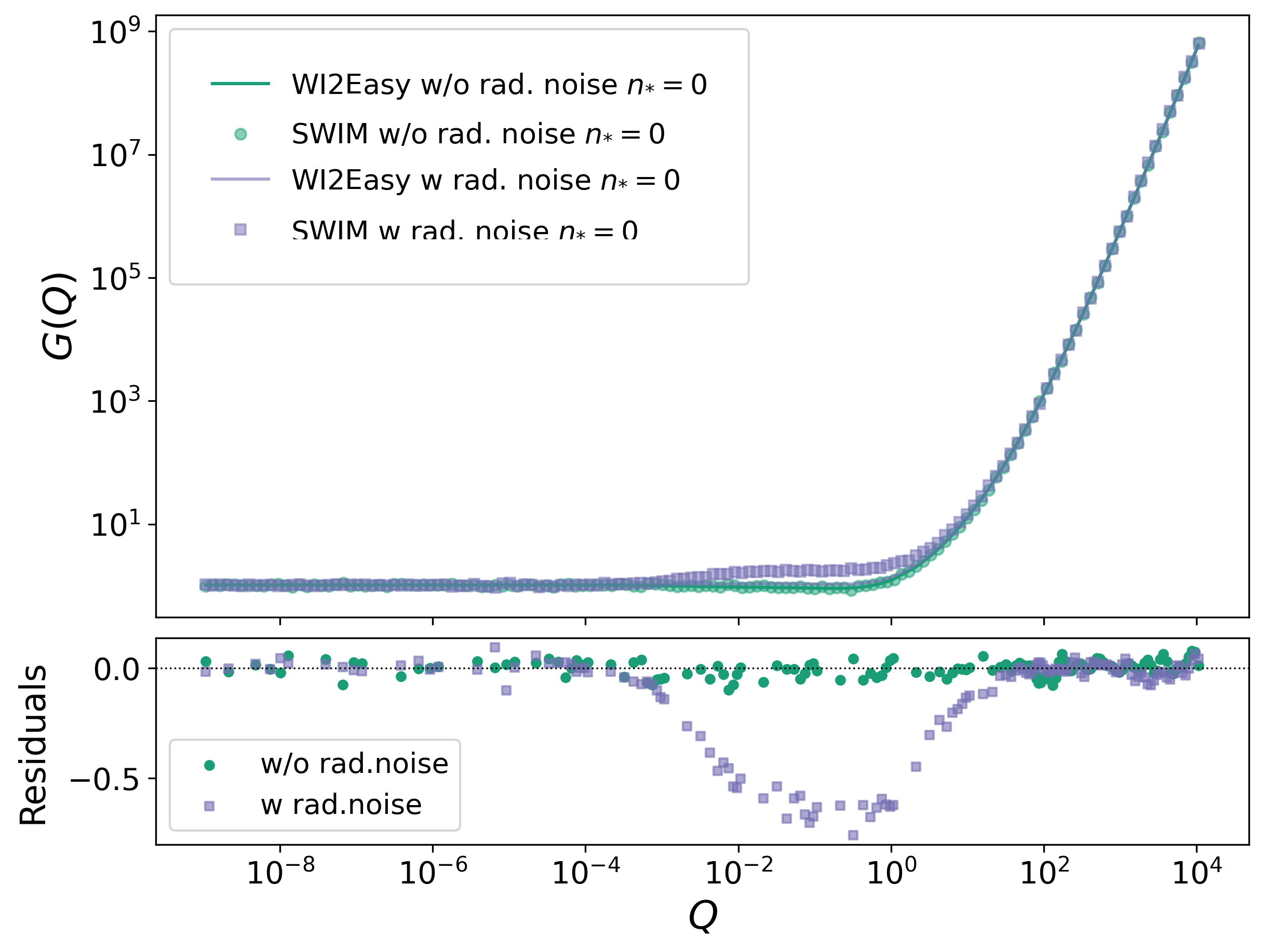}}
\subfigure[Fibre type I, $\Upsilon\propto T^3$]{\includegraphics[width=5cm]{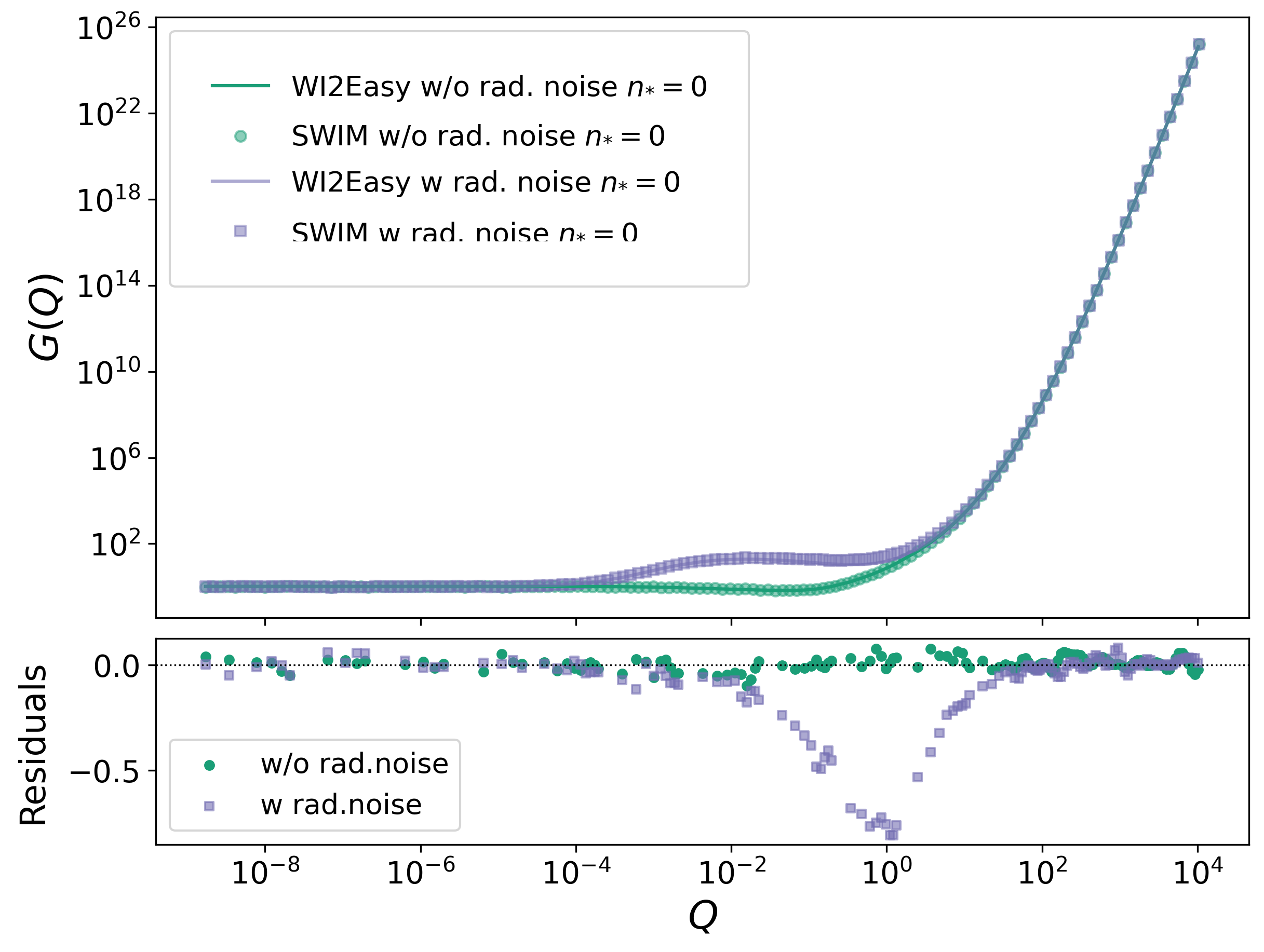}}
\subfigure[Cosine axion-type,  $\Upsilon\propto T$]{\includegraphics[width=5cm]{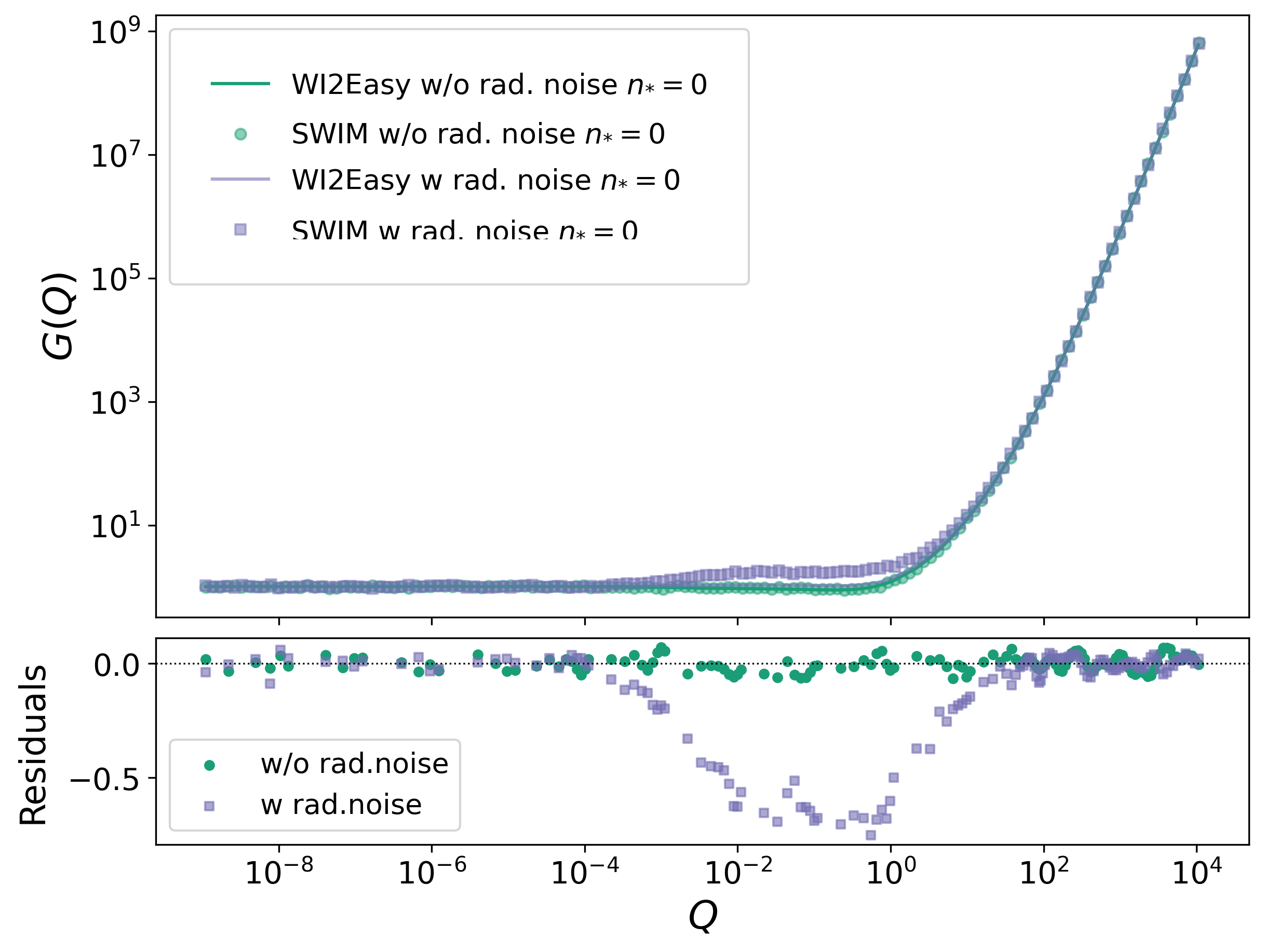}}\\
\subfigure[Cosine axion-type,  $\Upsilon\propto T^3$]{\includegraphics[width=5cm]{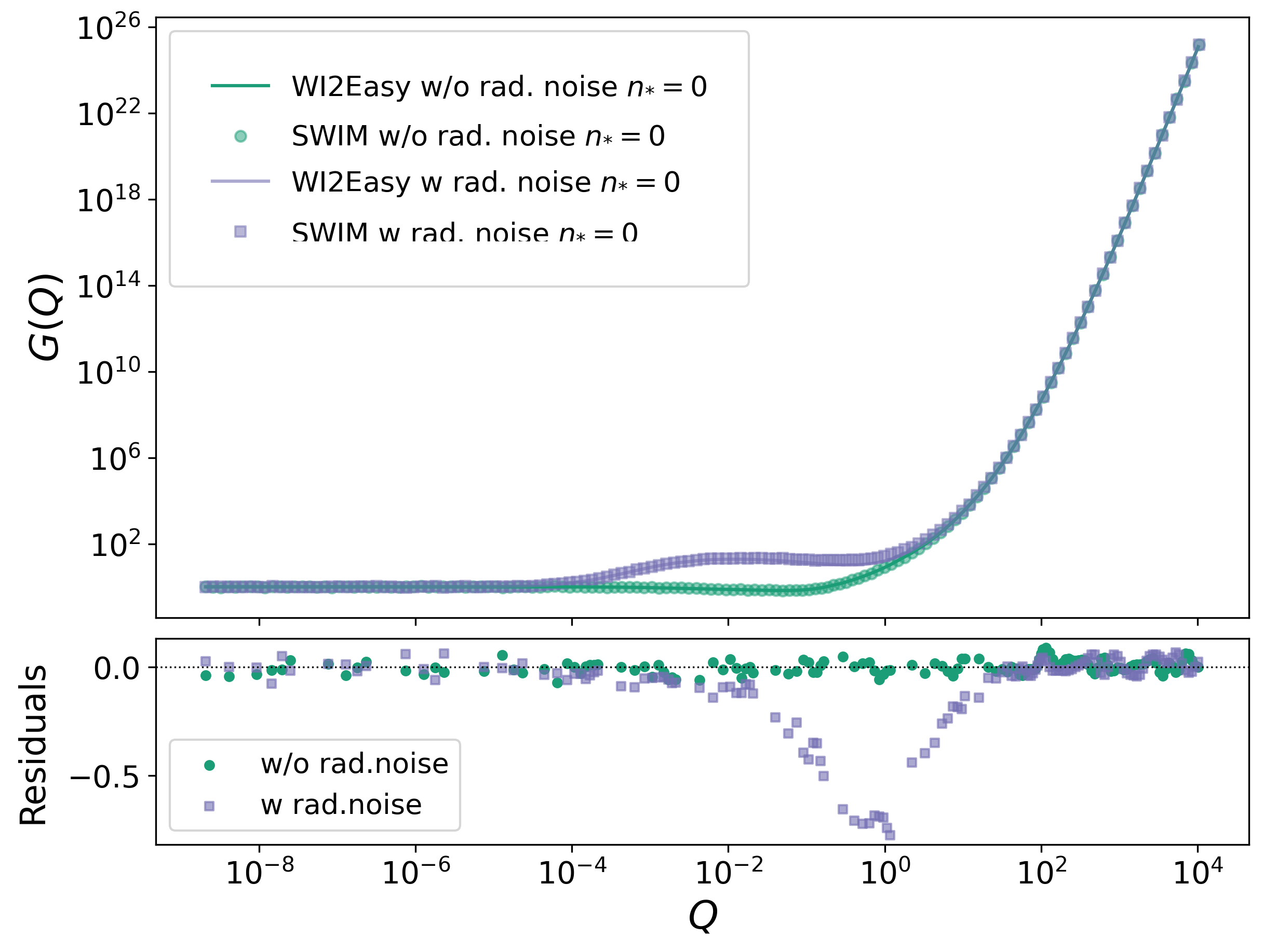}}
\subfigure[Quadratic Hiltop,  $\Upsilon\propto T$]{\includegraphics[width=5cm]{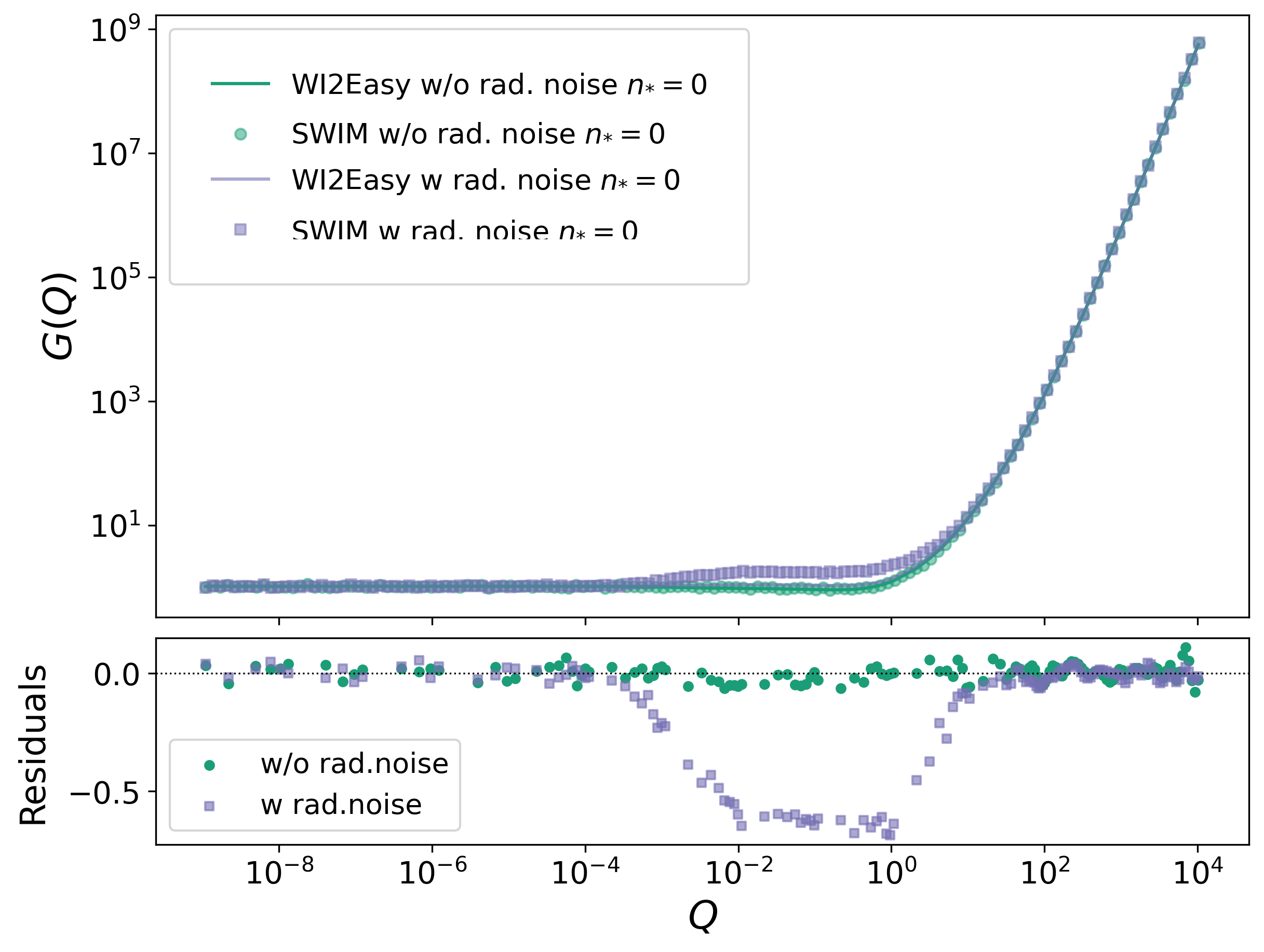}}
\subfigure[Quadratic Hiltop,  $\Upsilon\propto T^3$]{\includegraphics[width=5cm]{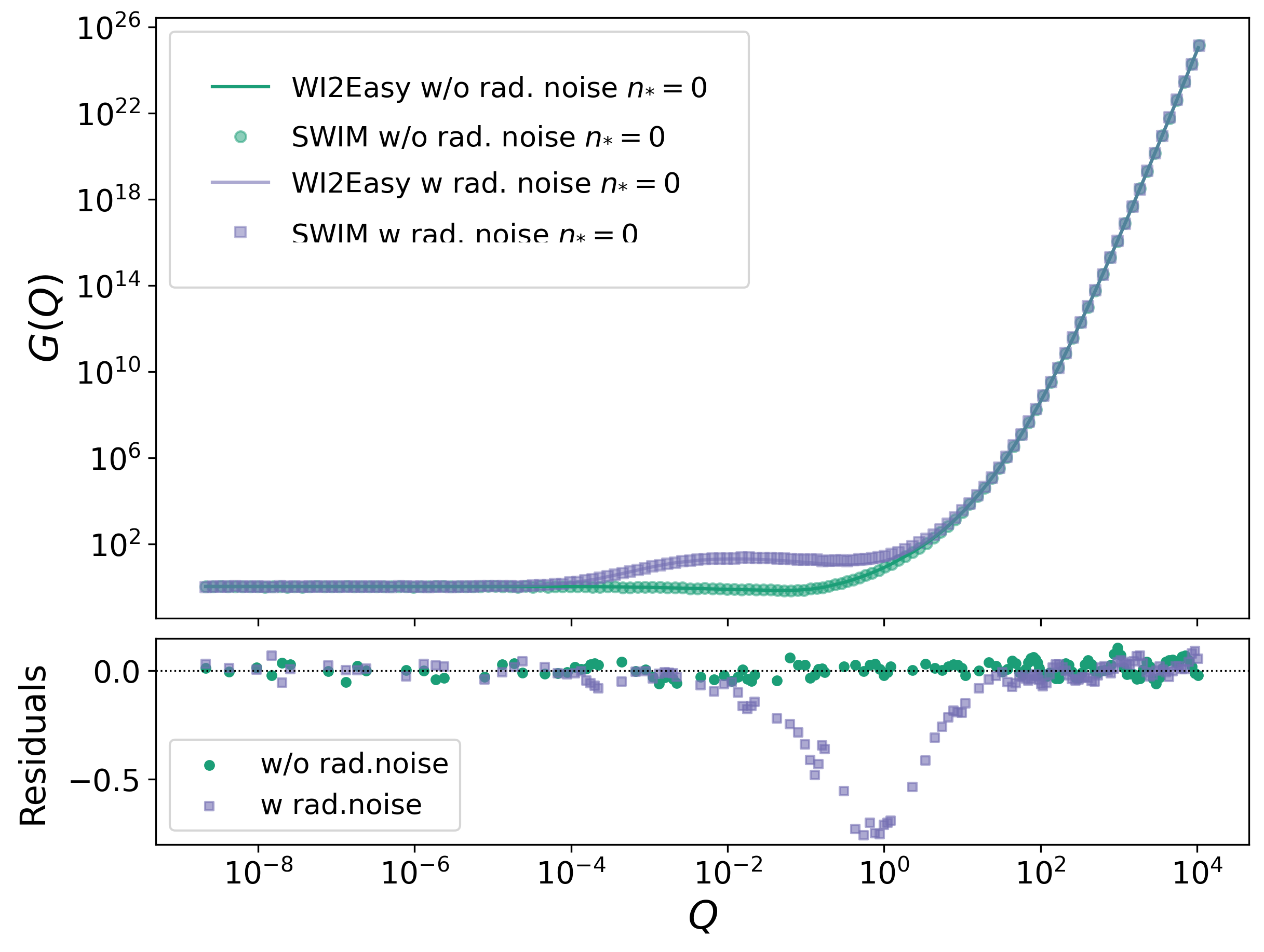}}\\
\subfigure[Hybrid inflation,  $\Upsilon\propto T$]{\includegraphics[width=5cm]{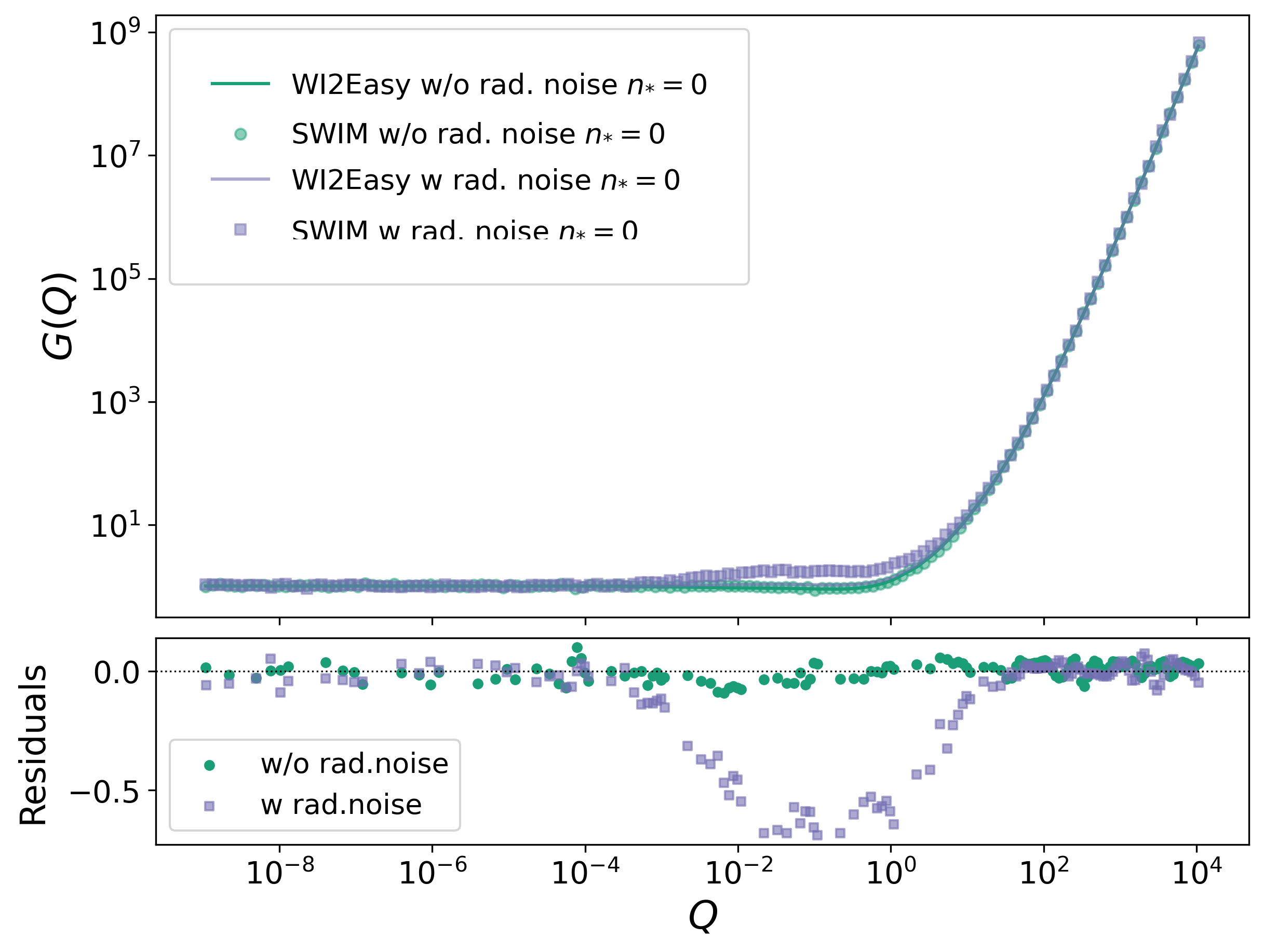}}
\subfigure[Hybrid inflation,  $\Upsilon\propto T^3$]{\includegraphics[width=5cm]{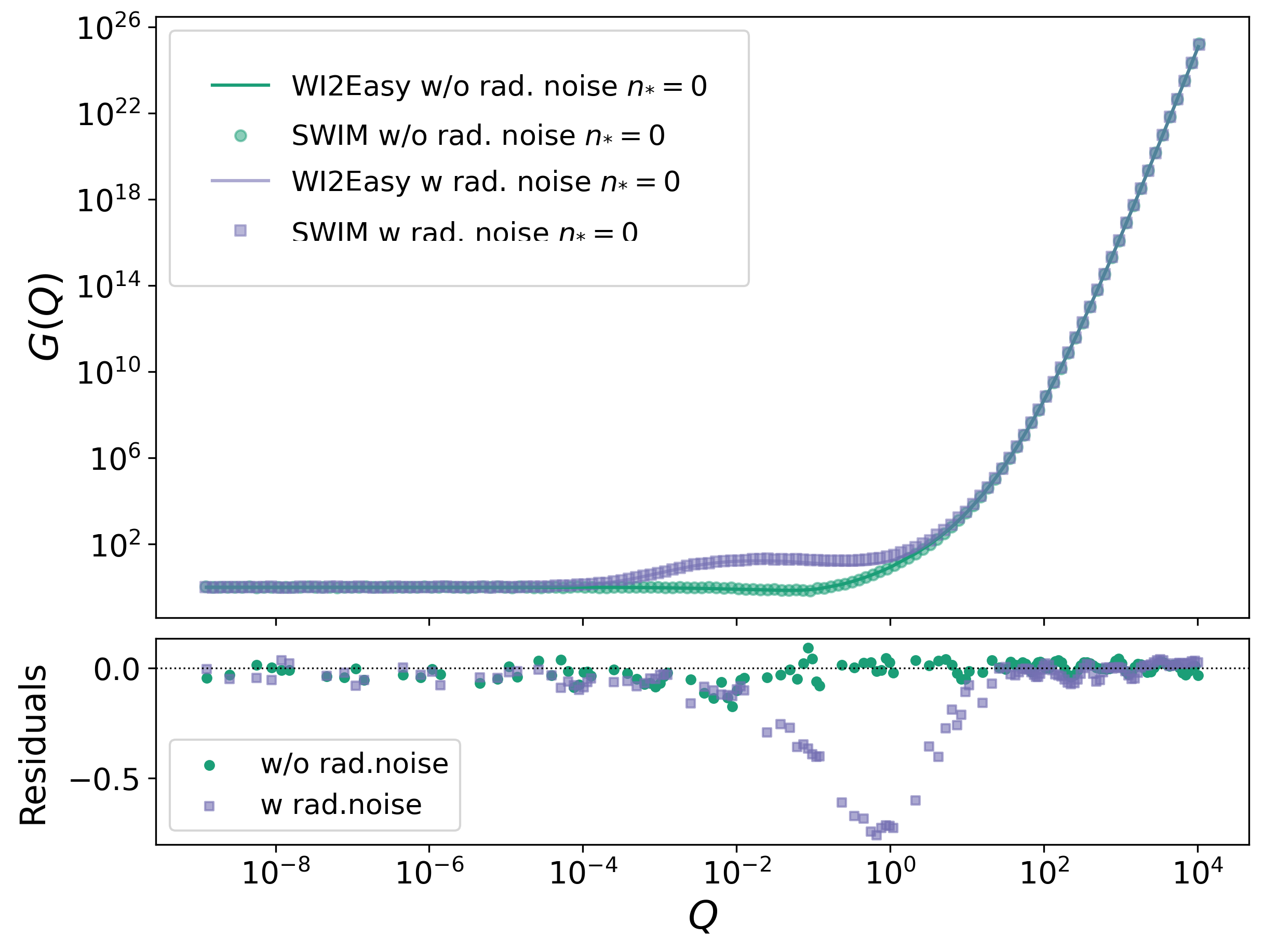}}
\subfigure[Quadratic monomial,  $\Upsilon_{\rm EFT}$]{\includegraphics[width=5cm]{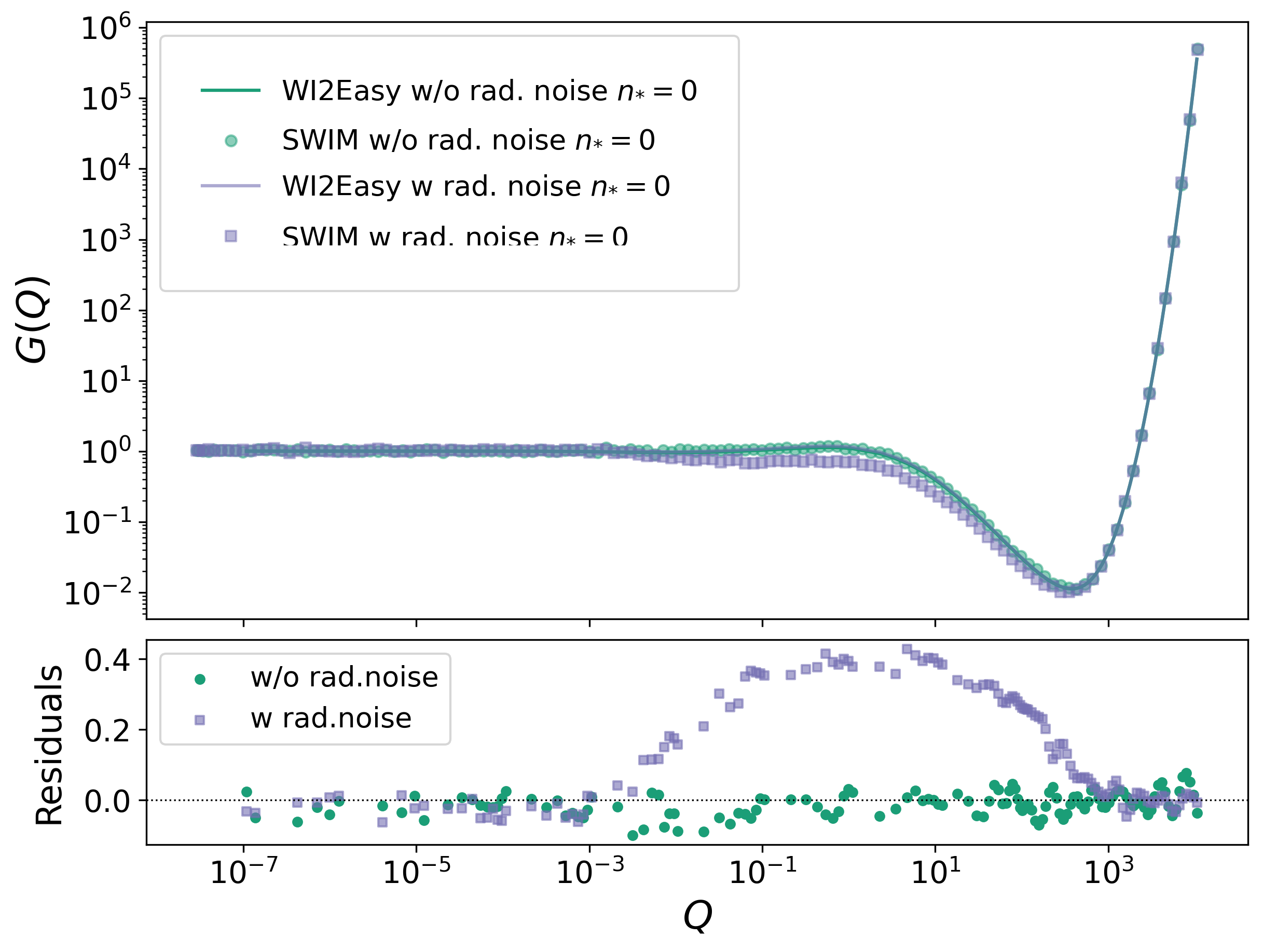}}
\caption{Comparing the outputs of \texttt{SWIM} with that of \texttt{WI2easy} for twelve represntative Warm Inflationary scenarios, while considering that the inflaton field has not thermalized with the thermal bath. The purple and green curves represent the cases when the noise term has been included in the radiation perturbation equation and when it is not, respectively. The lower panels in each plot shows the deviation of \texttt{SWIM} output from that of \texttt{WI2easy}.}
\label{GQ-noBE}
\end{figure}
\end{center}

\begin{center}
\begin{figure}[!htb]

\subfigure[Quartic monomial, $\Upsilon\propto T$]{\includegraphics[width=5cm]{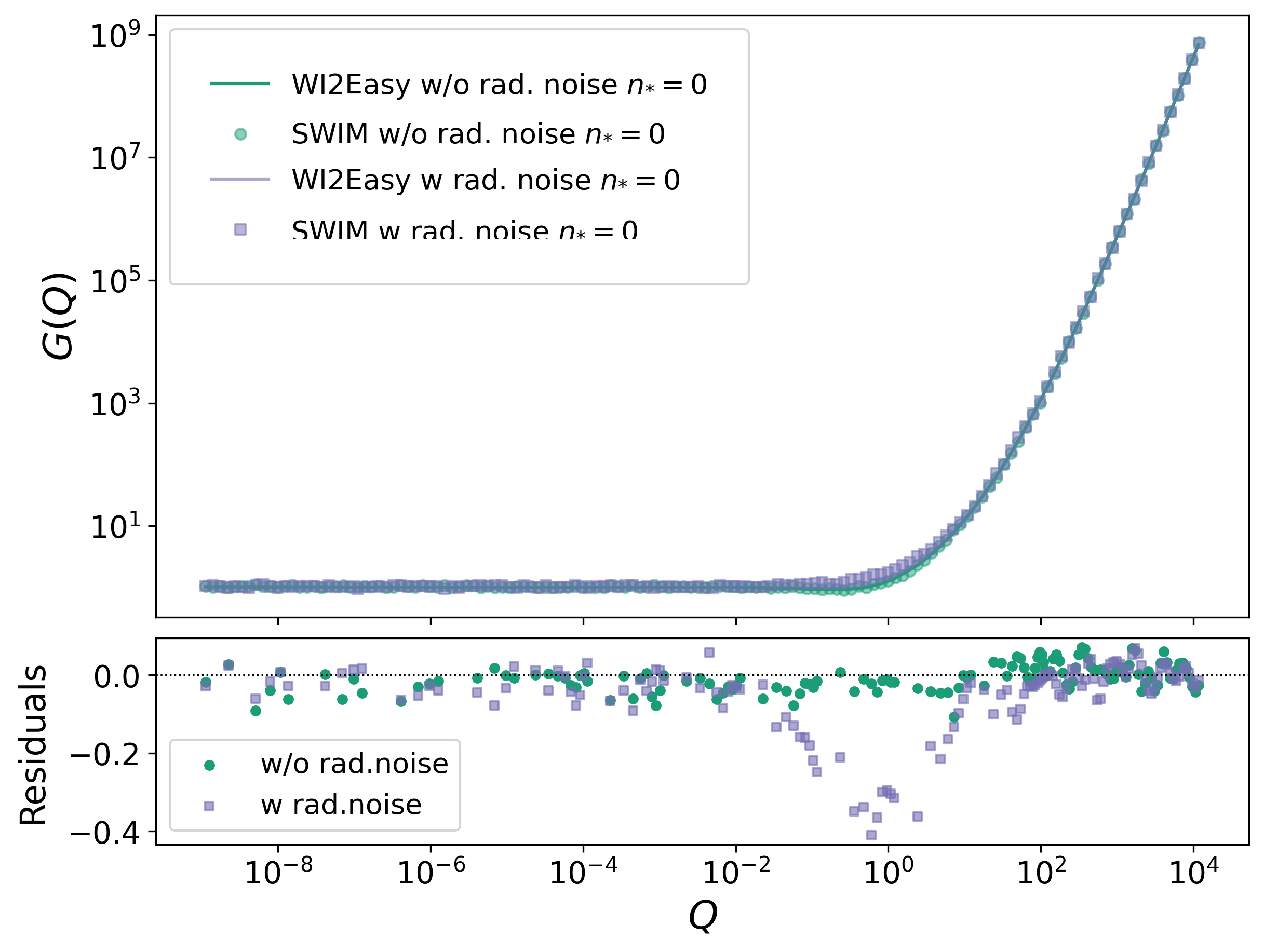}}
\subfigure[ Quartic monomial, $\Upsilon\propto T^3$]{\includegraphics[width=5cm]{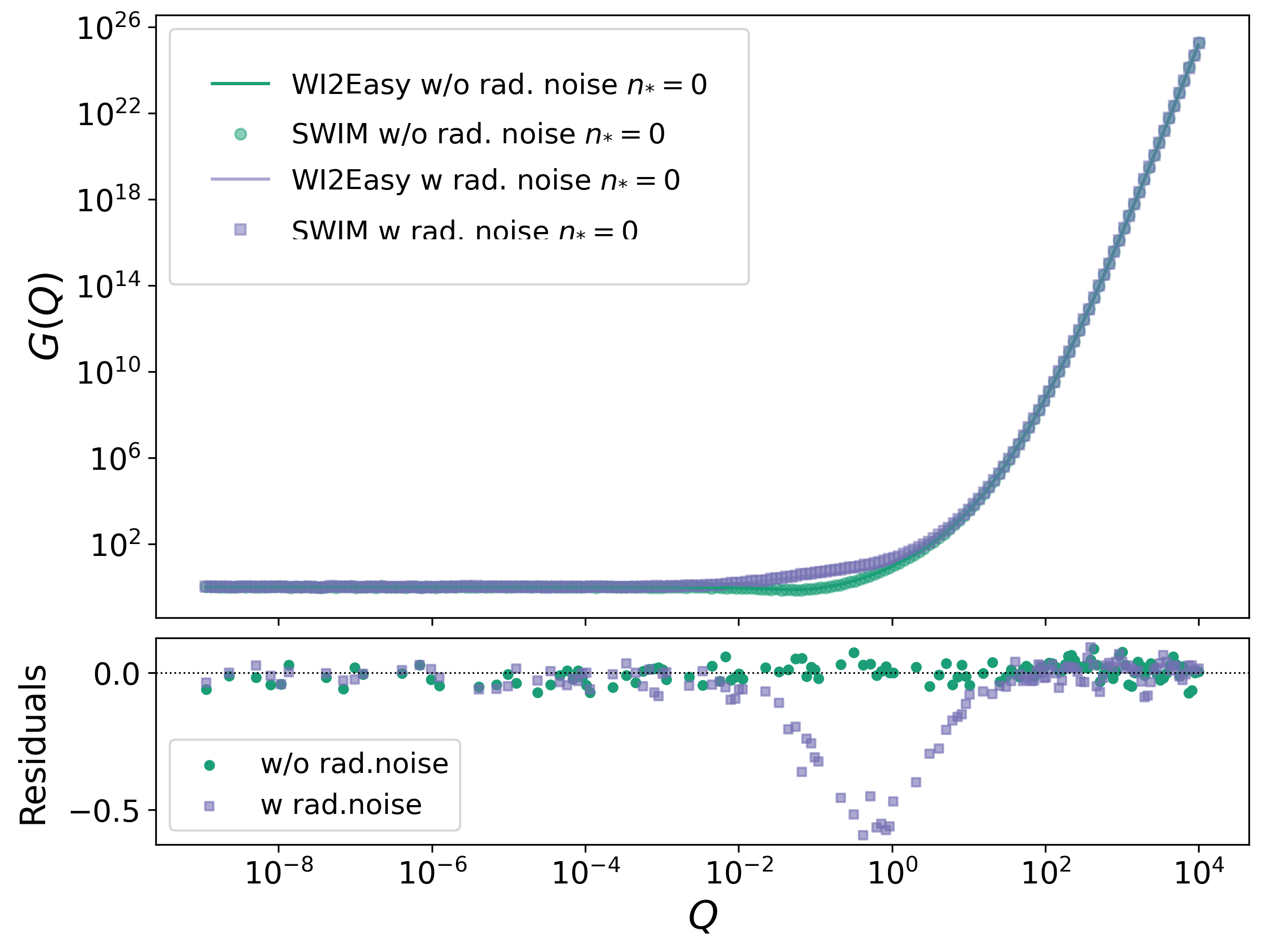}}
\subfigure[ Quadratic monomial, $\Upsilon_{\rm EFT}$ ]{\includegraphics[width=5cm]{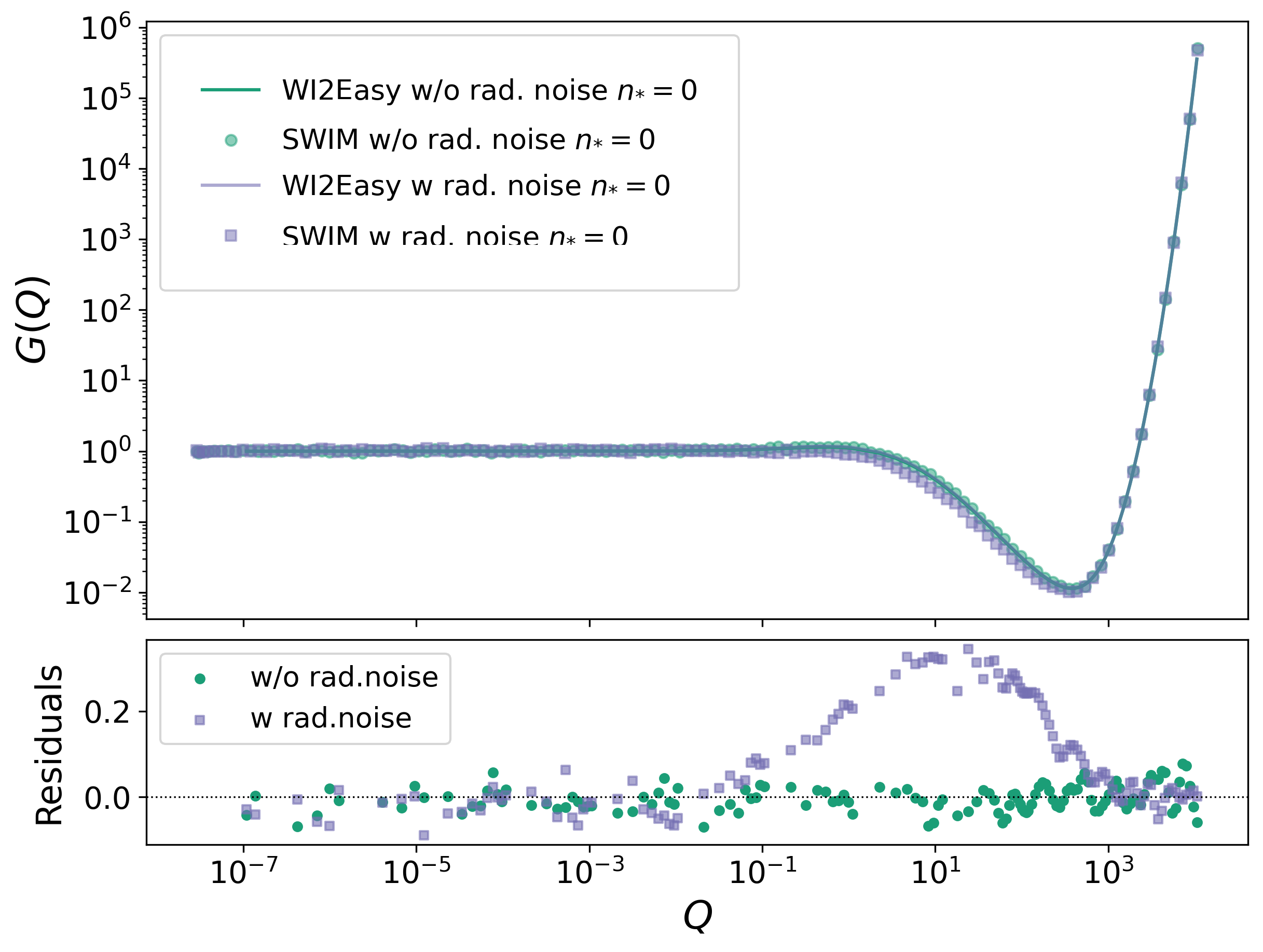}}

\caption{Comparing the outputs of \texttt{SWIM} with that of \texttt{WI2easy} for three representative Warm Inflationary scenarios, while considering that the inflaton field has thermalized with the thermal bath and has achieved a Bose-Einstein distribution. The purple and green curves represent the cases when the noise term has been included in the radiation perturbation equation and when it is not, respectively. The lower panels in each plot shows the deviation of \texttt{SWIM} output from that of \texttt{WI2easy}.}
\label{GQ-BE}
\end{figure}
\end{center}

Each of these cases can be studied with or without introducing the noise term in the radiation perturbation equation as well as considering whether the inflaton has thermalized or not. In Fig.~\ref{GQ-noBE}, we have studied twelve of such cases where the inflaton is not thermalized while comparing in each case the forms of $G(Q)$ with or without the radiation noise term (represented by the purple and green curves, respectively). We have included the Bose-Einstein distribution of the inflaton fluctuations considering the field to be thermalized in three representative cases illustrated in 
Fig.~\ref{GQ-BE}. Here too, we have compared the forms of $G(Q)$ with and without the radiation noise terms. All these plots are also compared with the outputs of \texttt{WI2easy} (the solid lines represent the outputs from \texttt{WI2easy} whereas the circle-squares represent the outputs from \texttt{SWIM}). Fig.~\ref{GQ-noBE} and Fig.~\ref{GQ-BE} confirm that \texttt{SWIM} is able to yield $G(Q)$ for a wide range of $Q$ values, $Q\in[10^{-8}, 10^4]$ for any given Warm Inflationary model with any potential and any form of dissipative coefficient. Its outputs also match with that of \texttt{WI2easy} in both weak $(Q\ll1)$ and strong $(Q\gg1)$ dissipative regimes. This confirms that the two mechanisms to determine WI power spectrum numerically, i.e., by solving stochastic perturbation equations or by solving the corresponding deterministic Fokker-Planck equation, are equivalent. However, we saw the output of \texttt{SWIM} to deviate from that of \texttt{WI2easy} in the moderate dissipative regime around $Q\sim1$ in all the cases studied here when one includes the noise term in the radiation perturbation equations. This discrepancy between the outputs of \texttt{SWIM} and \texttt{WI2easy} stems from the fact that \texttt{WI2easy} uses uncorrelated thermal noises (i.e., two independent $\xi_T$'s for inflaton and radiation perturbations equations) \cite{Rodrigues:2025neh}, whereas \texttt{SWIM} uses correlated thermal noises (i.e., same $\xi_T$ for inflaton and radiation perturbation equations) while solving for the perturbation equations at each step of the numerical evolutions. However, as the thermal noise terms $\xi_T$ in inflaton and radiation perturbation equations appear due to the energy-momentum conservation, they must be correlated, and hence \texttt{SWIM}'s output is more accurate than that of \texttt{WI2easy}. 

We have also compared the runtimes of \texttt{SWIM} and \texttt{WI2easy} for each of these models, and have observed that, despite solving several realizations of stochastic evolutions, \texttt{SWIM} is not only comparable to \texttt{WI2easy} but also outperforms \texttt{WI2easy} in most of the cases. We have furnished the performances of \texttt{SWIM} and \texttt{WI2easy} in Table~\ref{tab:runtime_table}.  It is to note that the longer initial condition search times in \texttt{SWIM} stem from the use of a robust brute-force optimization routine that repeatedly solves the exact background equations to determine the inflaton initial conditions yielding the desired number of $e$-folds. While this approach is computationally more expensive than dedicated root-finding methods adapted in \texttt{WI2easy}, it provides a model-independent framework applicable to arbitrary warm inflation potentials and dissipation coefficients. Furthermore, in the strong dissipative regime $(Q \gg 1)$, the background equations become increasingly stiff, requiring smaller adaptive integration steps and thereby increasing the computational cost. We aim to incorporate more efficient root-finding strategies to accelerate the initial condition search in the future versions of \texttt{SWIM}.

\begin{table}[h]
\centering
\resizebox{\textwidth}{!}{%
\begin{tabular}{|c|c|c|c|c|c|c|c|c|c|}
\hline
\multicolumn{2}{|c|}{\textbf{Model}} & \textbf{Parameters} & \textbf{$Q$ Range} & \textbf{Radiation} & \textbf{Thermalization} & \multicolumn{2}{|c|}{\textbf{Find ICs}} & \multicolumn{2}{|c|}{\textbf{Find $G(Q)$}}   \\
\multicolumn{2}{|c|}{}  &  &  & \textbf{Noise} & & \multicolumn{2}{|c|}{\textbf{Run-time(s)}}  & \multicolumn{2}{|c|}{\textbf{Run-time(s)}}   \\
\hline
$V(\phi)$ & $\Upsilon(\phi,T)$ & $N_e = 60$, $g_*=100$, & & & & WI2easy & SWIM & WI2easy & SWIM \\
 &  &  $V_0/M_{\text{Pl}}^4 = 10^{-14}$ & & & &  &  &  &  \\

\hline

\multirow{13}{*}{$\dfrac{V_0}{4}\phi^4$}
& \multirow{5}{*}{$C_{\Upsilon}\dfrac{T^3}{\phi^2}$}
& & $Q_{\rm min} = 10^{-9}$, & No & No & 205.779 & 5536 & 4491.23 & 1166.93\\
&  & & $Q_{\rm max} =  10^4$ &  &  &  &  &  & \\
&  & & & Yes & No & 207.867 & - & 4679.91 & 1164.35 \\
&  & & & No & Yes & 204.797 & - & 4088.3 & 1184.93 \\
&  & & & Yes & Yes & 211.18 & - & 4125.58 & 1185.23 \\

\cline{2-10}
& \multirow{4}{*}{$C_{\Upsilon}T^3$}
& & & No & No & 155.443 & 158.23 & 1381.73 & 569.22 \\
& & & & Yes & No & 157.139 & - & 1387.25 & 569.45 \\
& & & & No & Yes & 158.006 & - & 1285.41 & 589.19 \\
& & & & Yes & Yes & 157.204 & - & 1309.59 & 588.97 \\

\cline{2-10}
& \multirow{4}{*}{$C_{\Upsilon}T$}
& & & No & No & 193.603 & 631.54 & 2863.39 & 653.78 \\
& & & & Yes & No & 190.557 & - & 2870.6 & 650.28\\
& & & & No & Yes & 194.462 & - & 2612.57 & 680.27 \\
& & & & Yes & Yes & 193.383 & - & 2661.47 & 674.49 \\

\hline

\multirow{4}{*}{$V_0\left[F -4e^{-\frac{\phi}{\sqrt{3}}} + e^{-4\frac{\phi}{\sqrt{3}}} + R \left(e^{2\frac{\phi}{\sqrt{3}}}-1\right)\right]$}
& \multirow{2}{*}{$C_{\Upsilon}T^3$}
& $R = 3.05 \times 10^{-7}$, $F=3$ & & No & No & 174.241 & 225.1 & 957.086 & 646.19 \\
& & & & Yes & No & 69.8128 & - & 969.005 & 644.95\\

\cline{2-10}
& \multirow{2}{*}{$C_{\Upsilon}T$}
& & & No & No & 62.2239 & 290.62 & 1025.99 & 645.19 \\
& & & & Yes & No & 62.6676 & - & 1018.97 & 645.31 \\

\hline

\multirow{4}{*}{$V_0 \left[1+\cos\left(\dfrac{\phi}{f}\right)\right]$}
& \multirow{2}{*}{$C_{\Upsilon}T^3$}
& $f=5$ & & No & No & 353.145 & 687.14 & 847.748 & 582.36 \\
& & & & Yes & No & 358.585 & - & 847.118 & 583.77\\

\cline{2-10}
& \multirow{2}{*}{$C_{\Upsilon}T$}
& & & No & No & 300.877 & 879.56 & 966.601 & 589.08 \\
& & & & Yes & No & 296.587 & - & 1013.76 & 590.03 \\

\hline

\multirow{4}{*}{$V_0\left[1 - \gamma (\phi)^2\right]$}
& \multirow{2}{*}{$C_{\Upsilon}T^3$}
& $\gamma = 0.01$ & & No & No & 1698.96 & 556.09 & 2199.31 & 521.76 \\
& & & & Yes & No & 1720.21 & - & 2246.56 & 529.05\\

\cline{2-10}
& \multirow{2}{*}{$C_{\Upsilon}T$}
& & & No & No & 1864.67 & 503.79 & 950.84 & 503.46 \\
& & & & Yes & No & 1874.34 & - & 955.16 & 504.78 \\

\hline

\multirow{4}{*}{$\dfrac{M^4}{4\lambda} + \dfrac{V_0}{2}\phi^2$}
& \multirow{2}{*}{$C_{\Upsilon}T^3$}
& $M = 10^{-5}$, $\lambda = 0.1$ & & No & No & 104.11 & 559.974 & 814.914 & 563.24 \\
& & $g = 0.05$ & & Yes & No & 103.217 & - & 810.803 & 558.5\\

\cline{2-10}
& \multirow{2}{*}{$C_{\Upsilon}T$}
& & & No & No & 113.723 & 688.51 & 837.728 & 564.9 \\
& & & & Yes & No & 110.628 & - & 880.398 & 572.04 \\

\hline





\multirow{5}{*}{$\dfrac{V_0}{2}\phi^2$}
& \multirow{5}{*}{$C_\Upsilon g^4 \dfrac{\tilde M^2 T^2}{m_{\chi}^3} \left[1+ \dfrac{1}{\sqrt{2\pi}}\left(\dfrac{m_{\chi}}{T}\right)^{3/2}\right]e^{-m_{\chi}/T}$}
& $m_{\chi} = \sqrt{\dfrac{g^2\tilde M^2}{2} + \alpha^2 T^2}$ & $Q_{\rm min} = 10^{-7}$  & No & No & 69.4449 & 698.97 & 1018.27 & 683.98 \\
& & $\tilde M = 5.6 \times 10^{-5}$ & $Q_{\rm max} =  10^4$ & Yes & No & 69.7177 & - & 995.976 & 689.96\\
& & $g = 0.47$, $\alpha = \sqrt{1/8}$ &  & No & Yes & 69.4632 & - & 1018.57 & 723.82 \\
& & $g_* = 7.5$, $V_0 = 4.096 \times 10^{-13}$ &  & Yes & Yes & 68.6898 & - & 1055.22 & 716.87 \\




\hline
\end{tabular}%
}
\caption{Comparison of runtimes between \texttt{SWIM} and \texttt{WI2easy}. Specifications of the computer where these runtimes have been tested: CPU: AMD Ryzen 9 7900X 12-Core (24 threads), 
Memory: 16GB x 2 DDR5 6000MT/s,  OS: Fedora Linux 37,  Linux Kernel version: 6.5.12, Python version: 3.13, g++ (C++ compiler) version: 14.2.0, Mathematica Version: 13.2}
\label{tab:runtime_table}
\end{table}

\subsection{Model parameter estimation using the semi-analytical power spectrum module of \texttt{SWIM}}

A novel feature of \texttt{SWIM} is that one can directly use its output into \texttt{CAMB} and \texttt{Cobaya} to obtain best-fit values of the model parameters of the Warm Inflationary model one chooses to analyze using the methodology developed in \cite{Kumar:2024hju}. This feature is not available with the other two publicly available codes on Warm Inflation, namely \texttt{WarmSPy} and \texttt{WI2easy}. 

The semi-analytical power spectrum submodule of \texttt{SWIM} utilizes the $G(Q)$ factor computed numerically by the $G(Q)$ calculator submodule to evaluate the semi-analytical power spectrum numerically. Thus the user needs to first run the $G(Q)$ submodule of \texttt{SWIM} before using this submodule.The $G(Q)$ factor is read from the file \texttt{GQ\_Smooth.dat}, generated by the $G(Q)$ calculator submodule, and is internally interpolated to obtain a smooth representation of $G(Q)$ for use in the power-spectrum computation. However, this submodule of \texttt{SWIM} can also make use of the numerical $G(Q)$ generated by \texttt{WI2easy} or the analytical fit of $G(Q)$ obtained from \texttt{WarmSPy}. The \texttt{SA\_PS\_Calculator} directory of \texttt{SWIM} contains the file \texttt{model\_calc.cpp}, which the user needs to modify to specify the WI model under consideration. The directory also includes example input files for \texttt{Cobaya}, configured to perform parameter inference using current cosmological datasets. A schematic flowchart illustrating the workflow of this module is shown in Fig.~\ref{flowchart:Semi-Analytical_PS}. A guideline to use this submodule of \texttt{SWIM} is presented in Appendix~\ref{appendix:semi-analytical-ps}.


\begin{figure}
\centering

\includegraphics[width=1.0\textwidth]{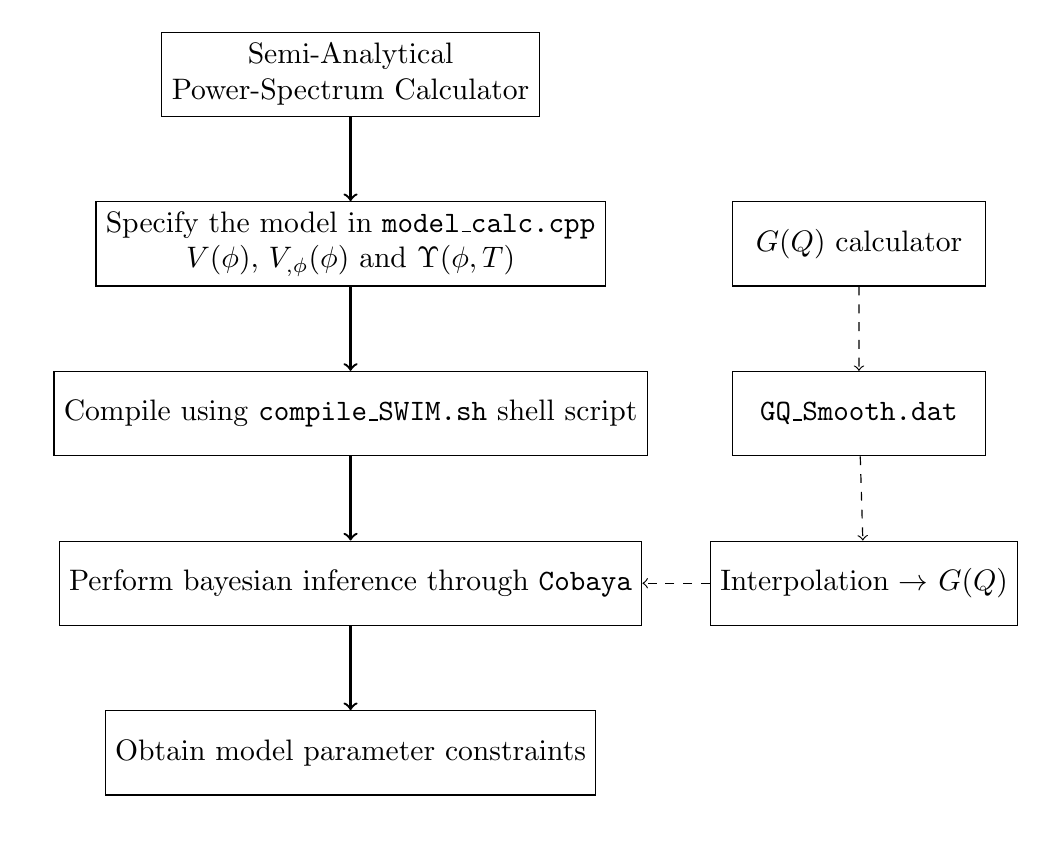}

\caption{Workflow of the semi-analytical power spectrum submodule of \texttt{SWIM}}
\label{flowchart:Semi-Analytical_PS}

\end{figure}

\section{Limitations of the semi-analytical approach and advantage of using \texttt{SWIM}}
\label{SWIM-module-3}

One major assumption that goes in determining the $G(Q)$ factor that appears in the semi-analytical scalar power spectrum of WI is that this correction factor depends solely on the parameter $Q$. However, as we demonstrate below with one concrete example, there can be cases where this correcting factor $G$ may depend on other model parameters, such as the overall normalization of the potential $V_0$ (defined in Eq.~(\ref{pot-form})), the relativistic degrees of freedom of the radiation bath $g_*$ (defined in Eq.~(\ref{rad-eqn})) or any other model parameters specific to the particle physics relization of the WI model. To be specific, the general understanding in the literature is that that the functional dependence of $G(Q)$ is sensitive to the temperature scaling of the dissipative coefficient $\Upsilon$, while exhibiting only a mild dependence on the detailed shape of the inflaton potential~\cite{Kamali:2023lzq}. This variation is important when one compares the form of $G(Q)$ across several WI models. However, what we noticed is that while fixing the functional forms of $\Upsilon$ and $V(\phi)$, i.e., within a single WI model, the form of $G(Q)$ varies significantly when one changes the model parameters, such as $V_0$ and $g_*$. In such cases, for each set of parameter values, represented by $\Theta$, $G(Q)$ can vary significantly. Therefore, constraining model parameters using the $G(Q)$ obtained from one particular set of parameter values $\Theta$, will instigate systematic biases in the inferred parameter ranges. 

\begin{figure}[h!]
    \centering
    
    \subfigure[Comparing $G(Q)$ with different values of $V_0$]{\includegraphics[width=7.5cm]{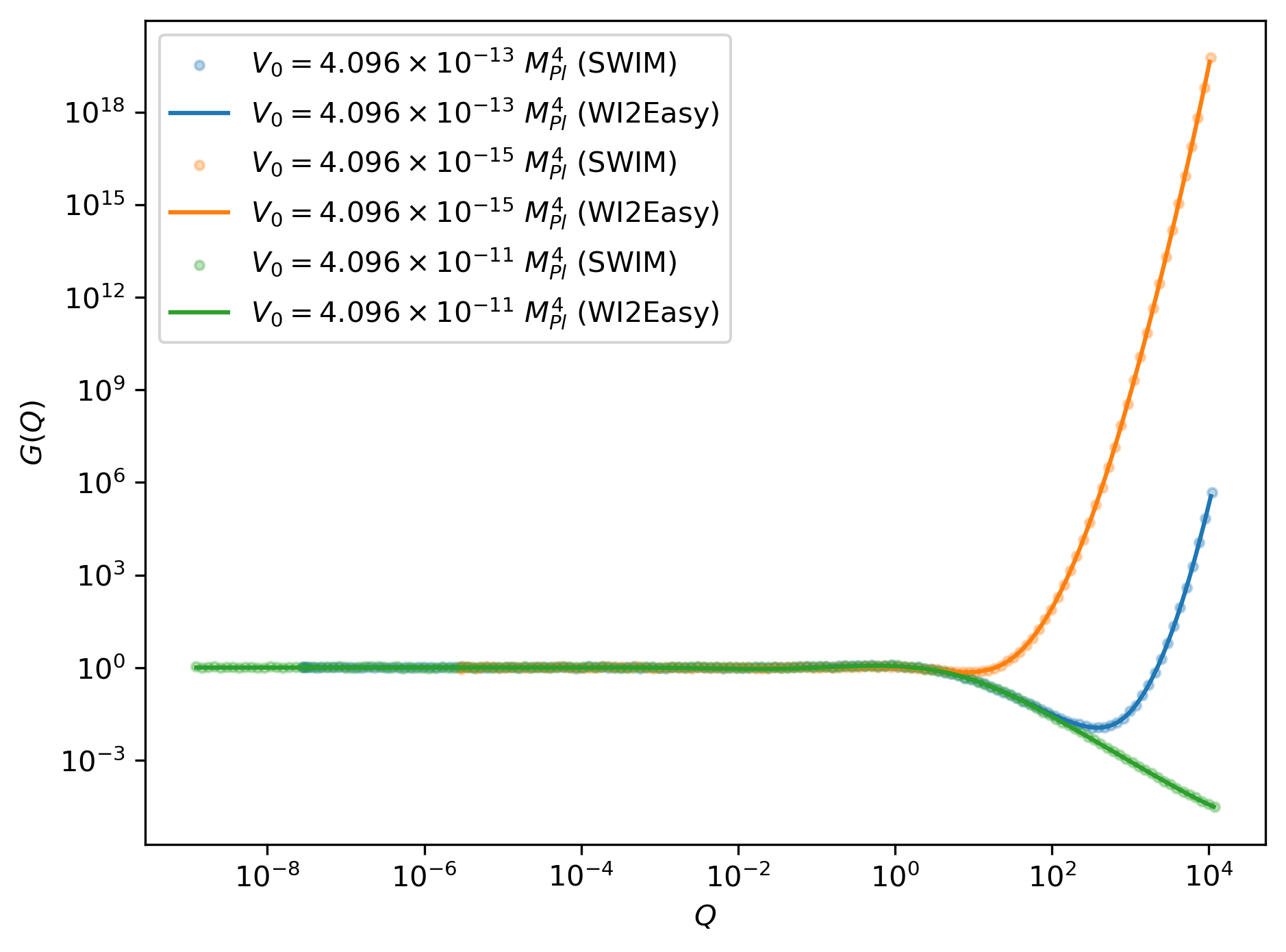}}
\subfigure[ Comparing $G(Q)$ with different values of $g_*$]{\includegraphics[width=7.5cm]{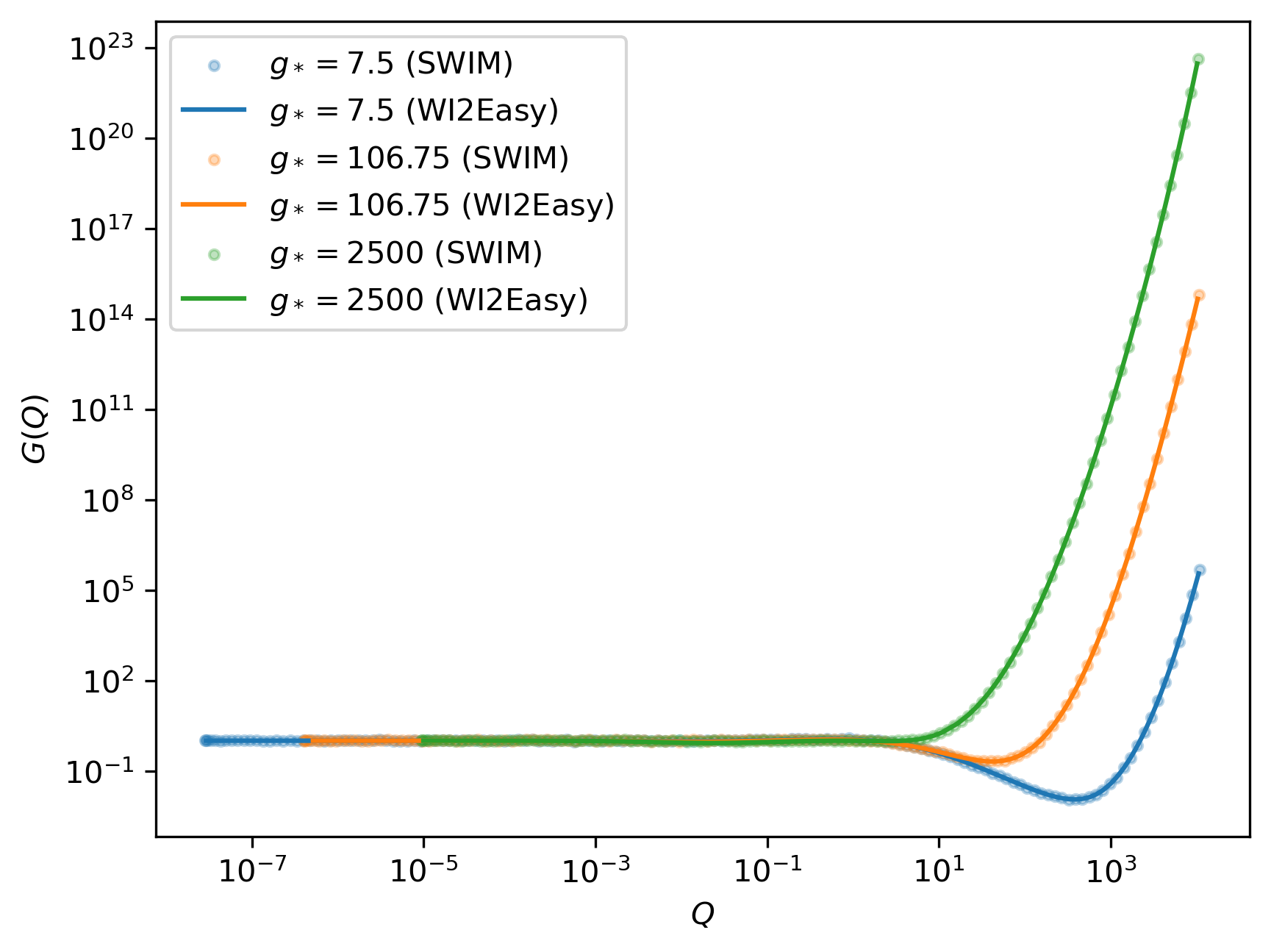}}  
   \caption{Comparison of $G(Q)$ calculated for the WI model presented in \cite{Bastero-Gil:2019gao} using both \texttt{SWIM} and \texttt{WI2easy}. The solid lines correspond to the results from \texttt{WI2easy} and the circular marks correspond to the results from \texttt{SWIM}. The inflaton perturbations are assumed to be not thermalized with radiation and we consider the case where thermal noise in the radiation perturbation equation is absent. Following~\cite{Bastero-Gil:2019gao}, we have set the other parameter values as
$g = 0.47$, $\alpha = \sqrt{1/8}$, and $\tilde{M} = 5.6 \times 10^{-5} M_{\text{Pl}}$.}
    \label{fig:EFT_Comparison}
\end{figure}

One such cases appears in the WI model studied in \cite{Bastero-Gil:2019gao}. In this model the dissipative coefficient is in the form as given in Eq.~(\ref{EFT-Ups}), and the potential is the quadratic one $V(\phi)=(V_0/2)(\phi/M_{\rm Pl})^2$. We find that the function $G(Q)$ exhibits a non-trivial dependence on the underlying model parameters, $V_0$ and $g_*$. The plots shown in Fig.~\ref{fig:EFT_Comparison}  illustrate how variations in $V_0$ and $g_*$ modify both the amplitude and shape of $G(Q)$ over the relevant range of $Q$. This residual parameter dependence implies that $G(Q)$ cannot, in general, be treated as a universal function of $Q$ alone. Consequently, any parameter inference procedure that relies on a pre-computed or fitted form of $G(Q)$ obtained for a fixed set of background parameters (as it is done in the case of the semi-analytical approach described in the previous section)  may lead to biased or inaccurate constraints.

To overcome this limitation, we propose to bypass the use of a pre-computed or fitted $G(Q)$ altogether. Instead, we perform parameter inference directly using the full numerical scalar power spectrum obtained from \texttt{SWIM}. In this approach, the coupled background and perturbation equations are solved self-consistently for each point in parameter space, and the resulting power spectrum is passed directly to the likelihood analysis. It is to note that neither \texttt{WarmSPy} nor \texttt{WI2easy} yields the numerical power spectrum as an output, while one can obtain the same using \texttt{SWIM}. This strategy eliminates the need for obtaining $G(Q)$ to calculate the Warm Inflationary power spectrum, and therefore avoids assumptions regarding the universality or parameter-independence of $G(Q)$. As a result, the inferred constraints fully incorporate the model-dependent dynamics of dissipation and radiation fluctuations, ensuring a consistent and robust exploration of the parameter space of any WI model.

In principle, one could directly interface the stochastic numerical perturbation solver with \texttt{Cobaya} to perform parameter inference. However, in practice, this approach is computationally prohibitive.  
To obtain a reliable scalar power spectrum, the solver must be run over multiple stochastic realizations, and the results averaged to reduce statistical noise. Furthermore, accurately computing the spectral index $n_s$ and its running $\alpha_s$ requires evaluating the power spectrum at a finite set of $k$-points spanning the observationally relevant range. Each of these points must, in turn, be averaged over multiple realizations.  When coupled with a MCMC sampler to explore the posterior distribution of model parameters, the computational cost increases dramatically. This is because the stochastic solver would need to be invoked many times per chain step, leading to a computational load that quickly becomes infeasible with typical available resources.

To make the bayesian parameter inference computationally practical we propose to use random forest regression (RFR) machine learning (ML) techniques \cite{Breiman2001, Conceicao:2023fjk} to build a surrogate model for the computationally expensive stochastic solver code. The RFR is model-independent and learns from data generated during the MCMC phase itself while simultaneously sampling the posterior and training the surrogate RFR model. The initial burn-in phase still relying on the true stochastic solver will be slow but as the RFR gets trained it quickly takes over, and the calls to the true solver are reduced that speeds up the MCMC process.

The key feature of this approach is that a reliability criterion derived from the random forest is used to dynamically decide whether to trust the surrogate prediction or to invoke the full numerical solver. In particular, the level of agreement among individual trees in the ensemble serves as an approximate indicator of prediction stability. For parameter points where the ensemble predictions are sufficiently consistent, the surrogate model is used directly, thereby reducing computational cost. Conversely, for points where the spread among tree predictions is large, indicating reduced reliability, the full numerical solver is employed to obtain an accurate power spectrum. This strategy ensures that the resulting posterior distribution remains as faithful as possible to that obtained using the full solver, while significantly improving computational efficiency.

The RFR emulator is trained \emph{on-the-fly} during the MCMC exploration phase, ensuring that it is optimized for regions of parameter space with high likelihood. Rather than learning the full numerical power spectrum as a function of $k$, the emulator is trained to map the model parameters (e.g., $V_0$, $g_*$, along with $\phi_i$ and $Q_i$) directly to the parameters of the fitted scalar power spectrum, $(A_s, n_s, \alpha_s, \beta_s)$ which is given in the form 
\begin{equation}\label{eq:fitting-ps}
     P_{\mathcal{R}}(k) = A_s \left(\frac{k}{k_P}\right)^{n_s-1 + \frac{\alpha_s}{2} \ln (k/k_P) + \frac{\beta_s}{6} \ln^2(k/k_P)},
\end{equation}
where $A_s$ is the scalar power amplitude, $n_s$ is the scalar spectral index, $\alpha_s$ and $\beta_s$ are the running and the running of the running of scalar spectral index, respectively, and $k_P$ denotes the pivot scale. 
This choice avoids learning the stochastic fluctuations present in the raw numerical spectrum and ensures that the emulator captures only the physically relevant, smoothed features of the power spectrum.
Internally, for each set of model parameters, the numerical solver produces a raw power spectrum, from which the parameters $(A_s, n_s, \alpha_s, \beta_s)$ are extracted using Eq.~\eqref{eq:fitting-ps}. These fitted parameters are then used both for likelihood evaluation and as training data for the emulator. The fitting procedure is essential for reducing stochastic noise in both the likelihood evaluation and the training data used by the RFR. While effective, random forest regression can struggle to capture smooth, continuous transitions because its predictions are built from many individual `if-then' splits. This can result in a jagged approximation of the data. Furthermore, the model's large memory footprint can make it difficult to deploy in environments where storage or speed is limited.

The emulator is implemented in the \texttt{Emulator/RF\_Acc\_Cobaya} subdirectory of \texttt{SWIM}, where example configurations are provided for performing parameter inference using \texttt{Cobaya}. Once sufficiently trained, the emulator provides an efficient surrogate for the numerical solver in the relevant region of parameter space, significantly reducing the computational cost. The trained emulator is specific to the WI model under consideration. Any change in the model, including changes in the treatment of perturbations (e.g., thermalization assumptions or the inclusion of radiation noise), requires the emulator to be retrained.

Moreover, as the numerical power spectrum is obtained by solving stochastic equations, the process itself incorporates some uncertainty in parameter estimation. Residual stochastic fluctuations in the numerical power spectrum persist even after smoothing, as repeated evaluations of the stochastic solver at fixed model parameters yield slightly different results. To quantify this effect, the script \texttt{std\_gen.py} performs repeated evaluations at identical input parameters, allowing an estimate of the uncertainty associated with the numerical solver. These fluctuations, arising from averaging over a finite number of realizations, introduce an intrinsic uncertainty in the computed spectrum, which can be incorporated into the likelihood as an additional variance. 

One can quantify the uncertainty arising due to the stochastic nature of the numerical power spectrum in the following way. Here, we incorporate observational constraints only on the scalar power amplitude \(A_s\) at the pivot scale \(k_p\), the spectral index at the pivot scale, \(n_s\), and the tensor-to-scalar ratio, \(r\). Based on these observables, one can define the Gaussian log-likelihood function as 
\begin{equation}\label{eq:log-likelihood}
   \ln \mathcal{L}(\Theta) = -\dfrac{1}{2} \sum_{i = A_s,n_s}\left[\ln\left(2\pi \sigma_i^2\right) + \left(\dfrac{x_{{\rm obs},i} - x_{{\rm model},i}(\Theta)}{\sigma_i}\right)^2\right] \, ,
\end{equation}
where \(\sigma_i\) denotes the observational uncertainty associated with each observable \(x_i\). The tensor-to-scalar ratio \(r\) is treated separately in the analysis, as only an upper-bound on $r$ is available. Here, we assume that \(A_s\) and \(n_s\) are uncorrelated for simplicity. The quantity \(x_{{\rm obs},i}\) denotes the measured mean value of each observable, while \(x_{{\rm model},i}(\Theta)\) represents the corresponding model prediction for a given set of parameters \(\Theta\).
Since the true  solver is intrinsically stochastic, its predictions can be modelled as draws from a Gaussian distribution with mean \(\bar{x}_{{\rm model},i}\) and standard deviation \(\sigma_{{\rm model},i}\). To account for this solver-induced uncertainty in the log-likelihood of Eq.~\eqref{eq:log-likelihood}, we empirically estimate \(\sigma_{{\rm model},i}\) by running the solver \(20\) times at the same parameter set \(\Theta\). In this work, we neglect any parameter or model dependence of \(\sigma_{{\rm model},i}\). The total uncertainty for each observable is then taken as
\begin{equation}\label{eq:total_uncertainty}
    \sigma_i^2 = \sigma_{{\rm obs},i}^2 + \sigma_{{\rm model},i}^2 \, , 
\end{equation}
where \(\sigma_{{\rm obs},i}\) is the observational uncertainty.

\begin{figure}
\centering

\includegraphics[width=1.0\textwidth]{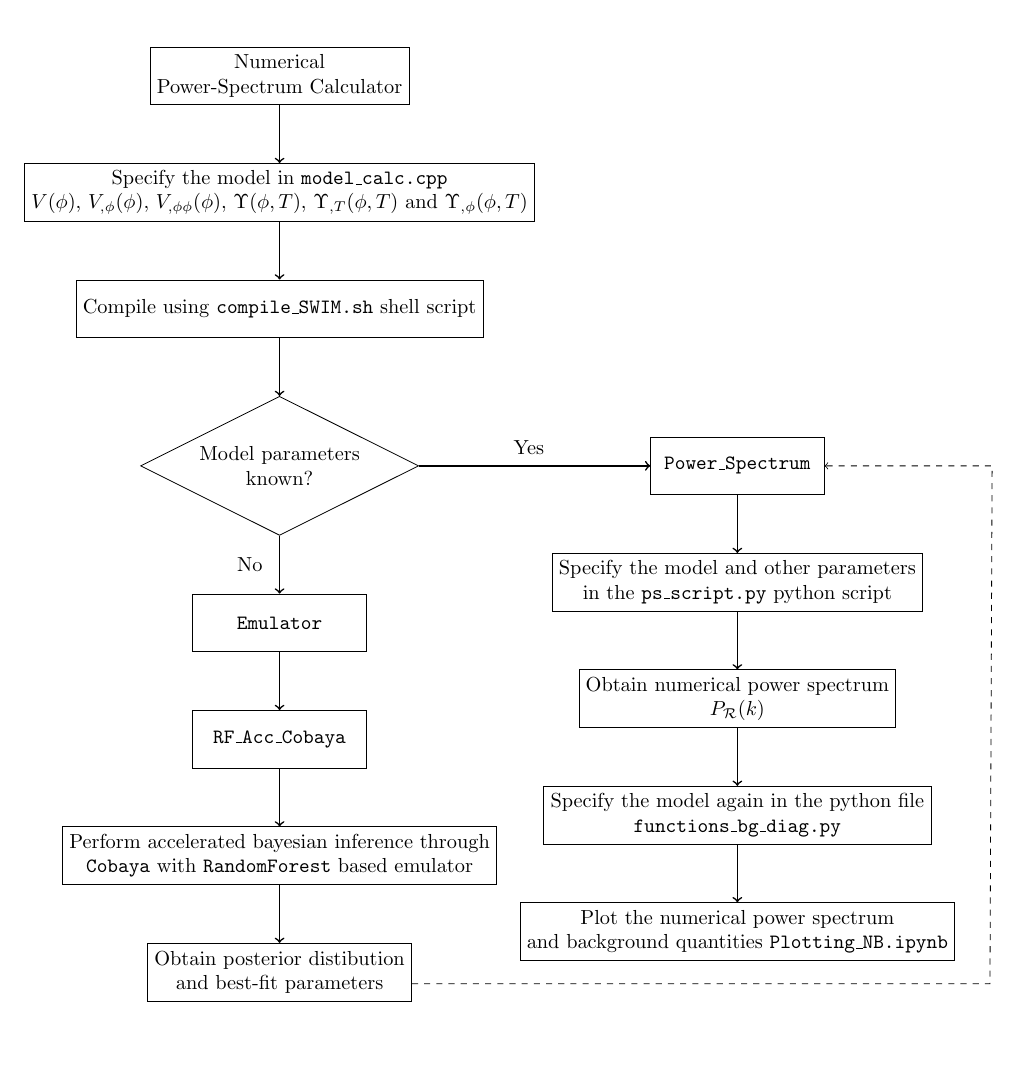}

\caption{Workflow of the numerical power spectrum submodule of \texttt{SWIM}}
\label{flowchart:Numerical_PS}

\end{figure}

We have depicted the workflow of the numerical power spectrum submodule of \texttt{SWIM} in Fig.~\ref{flowchart:Numerical_PS}. A step-by-step instructions on how to use this submodule of \texttt{SWIM} is given in Appendix~\ref{appendix:numerical-ps}.


\begin{figure}
    \centering
    \includegraphics[width=0.7\linewidth]{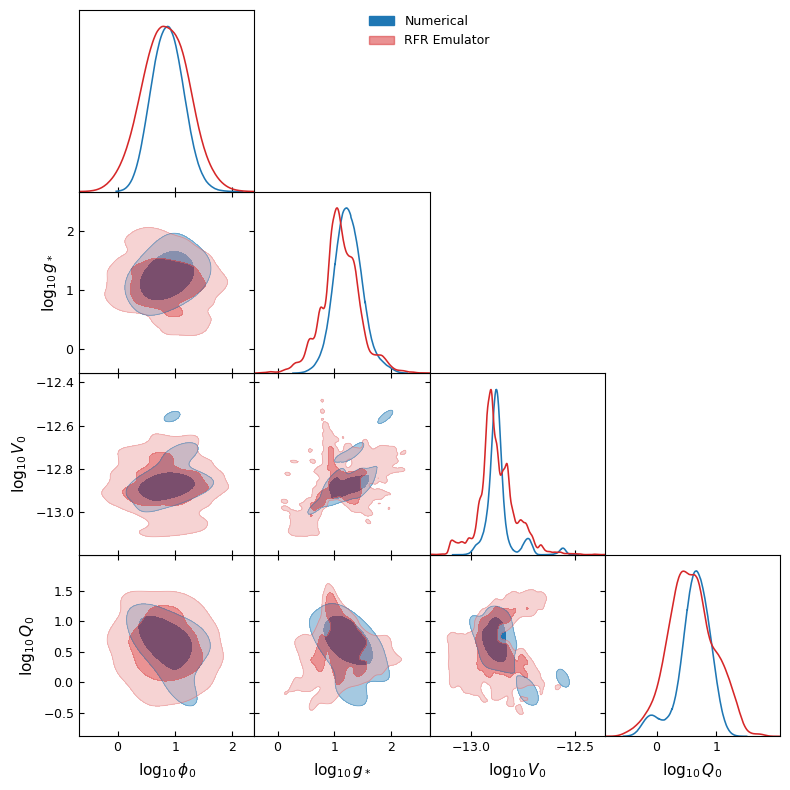}
    \caption{Comparison of posterior distributions for the WI model studied in \cite{Bastero-Gil:2019gao} obtained using the full numerical solver (blue) and the random forest emulator (red). The emulator reproduces the overall structure of the posterior. The small-scale irregularities in the emulator contours arise from the piecewise nature of the random forest approximation and the intrinsic stochastic variability of the numerical solver.}
    \label{fig:RFvNumerical_tri_plot}
\end{figure}

To put this methodology to test, we analyzed the WI model studied in \cite{Bastero-Gil:2019gao} in which, as shown previously, the $G(Q)$ factor depends on $V_0$ as well as $g_*$. We used the numerical power spectrum submodule of \texttt{SWIM}, and the posterior distribution of the model parameters obtained from \texttt{SWIM} are furnished in the triangle plot given in Fig.~\ref{fig:RFvNumerical_tri_plot}. In this figure, we have compared the posterior distributions of the model parameters obtained using the full numerical solver and the trained RFR emulator, and have shown that they are in good agreement, indicating that the emulator captures the relevant parameter dependencies. Small differences in the contours, including mild irregularities in the emulator results, arise from the piecewise nature of the random forest approximation and the intrinsic stochastic variability of the numerical solver. These effects do not significantly impact the inferred parameter constraints. We have also compared the best-fit values of the model parameters in Table~\ref{tab:best-fit_Quad_EFT} and the values of the derived parameters in Table~\ref{tab:best-fit_derived_Quad_EFT}  obtained from \texttt{SWIM} with the ones used in \cite{Bastero-Gil:2019gao}, and showed that they are all in good agreement. This guaranties the functionality of the numerical power-spectrum submodule of \texttt{SWIM}. It is to note that the RFR emulator employed in this work is trained adaptively during the sampling process rather than being pre-trained on a fixed dataset. Consequently, its predictive accuracy depends on the extent to which the relevant parameter space, particularly the high-likelihood regions, has been sampled. As the inference progresses and additional samples are accumulated, the emulator is iteratively refined and its agreement with the full stochastic numerical solver is expected to improve. The results presented in Table~\ref{tab:best-fit_Quad_EFT} and  Table~\ref{tab:best-fit_derived_Quad_EFT}  should therefore be regarded as representative examples and not necessarily as fully converged emulator runs. Nevertheless, the RFR approach provides a substantial computational advantage, achieving nearly three orders of magnitude speedup in likelihood evaluation relative to the full stochastic solver.

\begin{figure}
\centering

\includegraphics[width=0.6\textwidth]{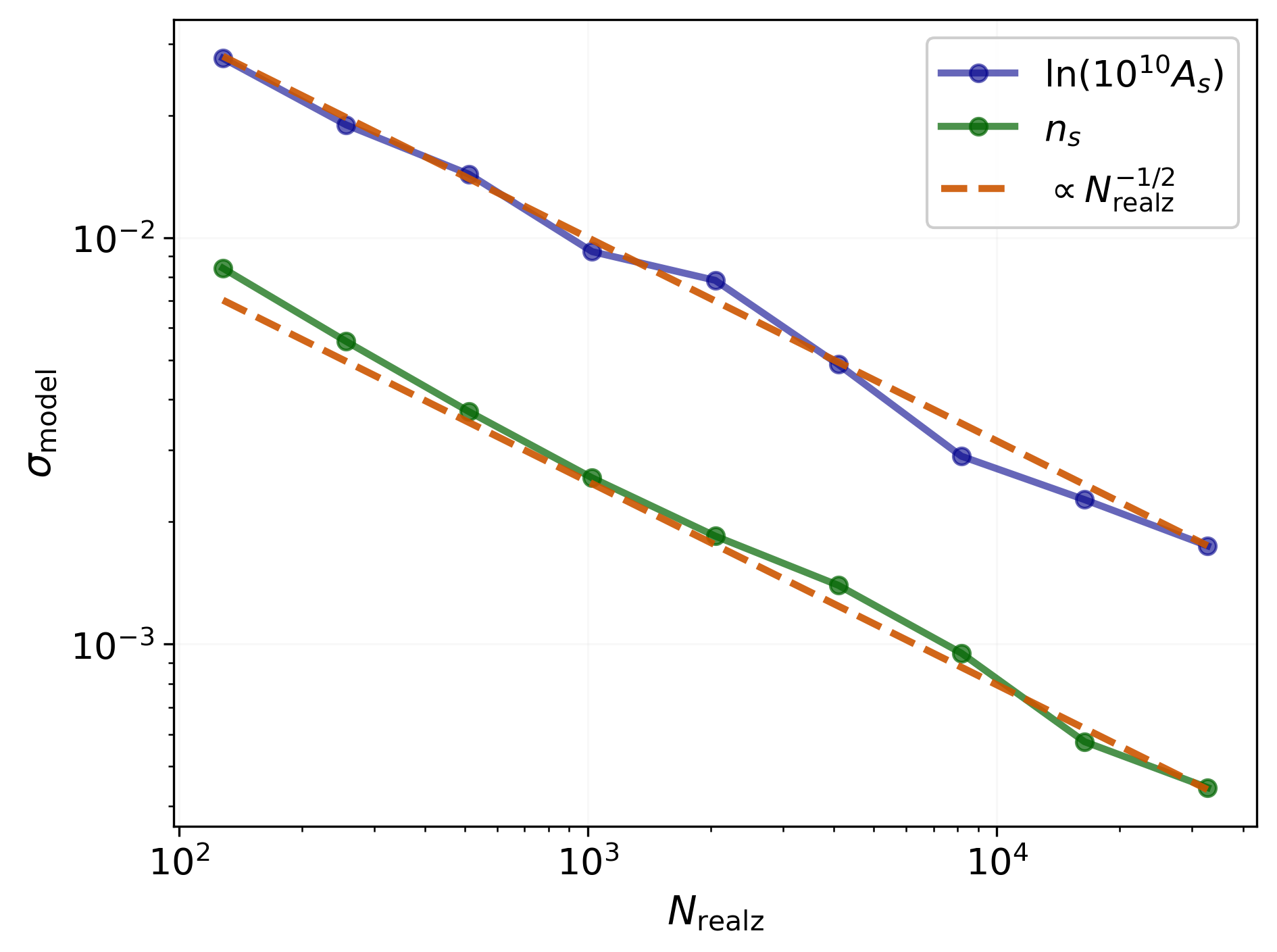}

\caption{Convergence of the stochastic perturbation solver with the number of realizations $N_\text{realz}$. The reference lines (orange-dashed) indicate a scaling $\sigma_{\rm model} \propto N_{\rm realz}^{-1/2}$.}
\label{sigma-convergence}

\end{figure}

 In order to quantify the stochastic uncertainty associated with a finite number of realizations, as estimated by the standard deviation $\sigma_{\rm model}$ defined in Eq.~(\ref{eq:total_uncertainty}), we repeatedly evaluated the perturbation solver at fixed WI model parameters and computed the standard deviation of derived inflationary observables $\ln(10^{10}A_s)$ and $n_s$. For a given value of $N_{\rm realz}$, the solver was executed 50 times, with each execution independently averaging over $N_{\rm realz}$ stochastic realizations. The standard deviation was then computed from these 50 samples. Fig.~\ref{sigma-convergence} shows the convergence of $\ln(10^{10}A_s)$ and $n_s$ with the number of realizations $N_{\rm realz}$. Fitting the numerical results yields slopes of $-0.511$ and $-0.527$ for $\ln(10^{10}A_s)$ and $n_s$, respectively, indicating that the observable uncertainties approximately follow an $N_{\rm realz}^{-1/2}$ dependence. 

On the other hand, a statistical error incorporated in the scalar power spectrum for the choice of number of realizations has been estimated in \cite{Ballesteros:2023dno} showing that to achieve $1\%$ accuracy one needs to choose at least $\mathcal{O}(10^4)$ realizations. We verified this estimation made in \cite{Ballesteros:2023dno} with our numerical analysis, and showed in Table~\ref{tab:stochastic_uncertainty} that they are in good agreement.

	

\begin{table}[t]
\centering
\begin{tabular}{l l l l l}
\hline
& $\phi_{i}/M_{\text{Pl}}$ & $Q_{i}$ & $V_0$  $(M_{	Pl}^4)$ & $g_*$ \\
\hline
WI model in \cite{Bastero-Gil:2019gao} & - & - & $4.096 \times 10^{-13}$ & $7.5$ \\
 Numerical Solver & $6.1856$ & $4.7371$ & $1.338 \times 10^{-13}$ & $13.7982$\\
 RFR emulator & $6.3541$ & $5.3533$ & $1.254 \times 10^{-13}$ & $16.4143$\\
\hline
\end{tabular}
\caption{Best-fit model parameters for the WI model studied in \cite{Bastero-Gil:2019gao} using the numerical solver (2048 realizations) and the RFR emulator. The corresponding Cobaya evaluation rates are $0.023\,\mathrm{s}^{-1}$ (approximately 44 s per likelihood evaluation) and $17.1\,\mathrm{s}^{-1}$ (approximately 0.06 s per evaluation), respectively, with the emulator providing a speedup of approximately $7.4\times10^2$ (nearly three orders of magnitude).}
\label{tab:best-fit_Quad_EFT}
\end{table}

\begin{table}[t]
\centering
\begin{tabular}{l l l l l l }
\hline
& $\phi_*/M_{\text{Pl}}$ & $Q_*$ & $A_s$ & $n_s$ & Observational  Dataset \\
\hline
WI model in \cite{Bastero-Gil:2019gao} & $1.0$ & $100$ & $2.2 \times 10^{-9}$  & $0.965$ & Planck\\
 Numerical Solver & $2.6857$ & $12.563$ & $2.22 \times 10^{-9}$ & $0.97$ & SPT+Planck+ACT+DESI\\
 RFR emulator & $2.3346$ & $17.2288$ & $2.057 \times 10^{-9}$ & $0.9707$ & SPT+Planck+ACT+DESI\\
\hline
\end{tabular}
\caption{Derived parameters corresponding to the best-fit model parameters for the WI model studied in \cite{Bastero-Gil:2019gao}, obtained using the full numerical solver and the RFR emulator. The RFR emulator directly predicts only the power-spectrum parameters defined in Eq.~\eqref{eq:fitting-ps} and does not provide information on the background evolution, such as $(\phi_*, Q_*)$. To report these quantities for the emulator case, the best-fit model parameters inferred using RFR are passed through the numerical solver to compute the corresponding background values.}
\label{tab:best-fit_derived_Quad_EFT}
\end{table}

\begin{table}[t]
	\centering
	\begin{tabular}{ccc}
		\hline
		$N_{\rm realz}$ & \texttt{SWIM} & Uncertainties estimated in \cite{Ballesteros:2023dno} \\
		\hline
		$10^3$ & $4.3\%$ & $3.0\%$ \\
		$10^4$ & $1.4\%$ & $1.0\%$ \\
		$10^5$ & $0.4\%$ & $0.3\%$ \\
		\hline
	\end{tabular}
	\caption{Relative stochastic uncertainty in estimating $A_s$ for different choices of number of realizations}
	\label{tab:stochastic_uncertainty}
\end{table}


\section{Discussion and Conclusion}
\label{summary}

Determination of the inflationary scalar power spectrum numerically is unavoidable in Warm Inflation, as analytically one can only obtain a power spectrum with significant approximations \cite{Graham:2009bf, Bastero-Gil:2011rva}. One of the main approximations that goes into determining the approximated analytical form of the scalar power spectrum in WI is to assume that the dissipative coefficient doesn't depend on the temperature $T$ of the radiation bath which helps the inflaton perturbation equation decouple from the radiation perturbation equation rendering it suitable to solve analytically \cite{Graham:2009bf, Ramos:2013nsa}. Though an approximate analytical form of the scalar power spectrum can be obtained, one then needs to multiply it with a correcting factor $G(Q)$ that accounts for all the approximations made during evaluating the approximated analytical form. The $G(Q)$ factor is essentially the ratio of the numerical power spectrum, which is the true power spectrum obtained numerically without any approximations made, and the approximated analytical power spectrum as defined in Eq.~(\ref{GQ}) \cite{Kamali:2023lzq}. Two publicly available codes, namely \texttt{WarmSPy}  (written in \texttt{Python}) \cite{Montefalcone:2023pvh} and \texttt{WI2easy} (written in \texttt{Mathematica}) \cite{Rodrigues:2025neh}, already exist in the literature that can be used to generate the $G(Q)$ factor appearing in the Warm Inflationary scalar power spectrum.  Though \texttt{WarmSPy} solves directly the Warm Inflationary stochastic perturbation equations,  this code is quite restrictive as it can  handle only a handful of inflationary potentials and specific forms of dissipative coefficients, and also is not quite accurate as it assumes $Q$ to be constant in its simulations. On the other hand, \texttt{WI2easy} solves the deterministic Fokker-Planck equations of the probability distributions of the perturbed quantities of WI to determine $G(Q)$, can accommodate any form of inflationary potentials and dissipative coefficients, and also solves the true background equations with evolving $Q$. 

However, even after obtaining the semi-analytical scalar power spectrum with the correcting $G(Q)$ factor, one needs to feed this power spectrum numerically to the MCMC codes, like \texttt{Cobaya} or \texttt{COSMOMC}, if one wishes to constrain the model parameters with the current CMB data. A generalised methodology to do so has already been developed in \cite{Kumar:2024hju}. None of the two publicly available codes, \texttt{WarmSPy} and \texttt{WI2easy}, is designed to feed the generated power spectrum directly to these MCMC codes, and thus cannot be used directly to constrain model parameters in WI. 

\texttt{SWIM} (written in \texttt{C++} and \texttt{Python}) can not only generate the $G(Q)$ factor for any Warm Inflationary model, i.e., for any inflationary potential and any form of dissipative coefficient, but also is integrated with \texttt{Cobaya} \cite{Torrado:2020dgo} that enables the user to perform parameter estimation of any WI model given the current CMB data. \texttt{SWIM}, unlike \texttt{WI2easy}, solves the full stochastic perturbations equations of WI to generate the numerical power spectrum. We have compared the $G(Q)$'s generated by \texttt{SWIM} with that of \texttt{WI2easy} for several Warm Inflationary models in Sec.~\ref{GQ-subsection} and have showed that they match perfectly in the weak ($Q\ll1$) as well as strong ($Q\gg1$) dissipative regime but differ slightly in the moderate dissipative regime ($Q\sim1$) when the noise term is included in the radiation perturbation equations. This not only ensures the functionality of \texttt{SWIM} but also verifies the claim made in \cite{Ballesteros:2023dno} that the Fokker-Planck approach of obtaining the WI power spectrum is equivalent to the standard stochastic formalism in the weak and strong dissipative regimes. But our analysis also show that \texttt{WI2easy}'s $G(Q)$ deviates from the one generated by \texttt{SWIM} when one explores the moderate dissipative regimes with the noise term included in the radiation equation. This discrepancy results from the fact that \texttt{WI2easy} uses uncorrelated thermal noise (in contrast to correlated thermal noise used in \texttt{SWIM}) while solving the perturbation equations. As the thermal noise term appears in both the inflaton as well as the radiation perturbation equations due to the conservation of the energy-momentum tensor, they must be correlated, and hence \texttt{SWIM}'s output is more trustworthy than that of \texttt{WI2easy} in these moderate dissipative regimes.  Moreover, even though \texttt{SWIM} solves the stochastic perturbation equations that needs to be performed and averaged over several realisations while \texttt{WI2easy} essentially solves deterministic Fokker-Planck equations where single realisation is sufficient, we showed in Table~\ref{tab:runtime_table} that in most cases \texttt{SWIM} outperforms \texttt{WI2easy} as far as the runtimes are concerned. 

The semi-analytical approach of obtaining the scalar power spectrum in WI falls short in cases where the factor $G(Q)$ depends significantly on other model parameters, such as the overall normalization of the potential $V_0$, the relativistic degrees of freedom $g_*$ of the radiation bath, or any other parameters specific to the particle physics construction of that WI model. The semi-analytical approach are insufficient in such cases as in the semi-analytical approach it is assumed that the correcting factor $G(Q)$ depends solely on  $Q$ and not on any other parameters. We demonstrated one such case where such a scenario occurs i.e., the WI model studied in \cite{Bastero-Gil:2019gao}, where $G(Q)$ varies significantly with $V_0$ and $g_*$. In such cases, if one wishes to constrain the model parameters, then the standard semi-analytical approach is not suitable, and one then needs to feed in the full numerical power spectrum in the MCMC codes. Neither \texttt{WarmSPy} nor \texttt{WI2easy} yields the numerical power spectrum as an output. However, by construction, the user can get the numerical power spectrum as an output in \texttt{SWIM}. In the numerical power spectrum submodule of \texttt{SWIM} we have specifically bypassed generating the $G(Q)$ factor, and have integrated \texttt{SWIM} with \texttt{Cobaya} so that the numerical power spectrum can directly be fed for the MCMC analysis. To reduce the computational cost, we implement the Random Forest regression Machine Learning techniques that enables the code to get trained from the initial MCMC analysis and once the RFR is trained, it quickly takes over the job of parameter estimation reducing the cost of rigorous MCMC analysis. We analysed the model presented in \cite{Bastero-Gil:2019gao} with this methodology using \texttt{SWIM}, and showed that they are in good agreement which ensures the functionality of this submodule of \texttt{SWIM}. However, this submodule of \texttt{SWIM} is not specific to cases where $G(Q)$ has dependancies on other model parameters. Anyone who wishes to constrain the model parameters of any WI model without generating the $G(Q)$ factor, can directly use this submodule to serve the purpose, though this process will always be computationally more costly than using the semi-analytical power spectrum approach. 

In summary, \texttt{SWIM} is so far the only single numerical code for WI which a user can use to perform a complete analysis of their preferred WI model, starting from obtaining the form of the scalar power spectrum to constraint it with the current CMB data. 

\appendix
\section{Instructions to install \texttt{SWIM}} 
\label{appendix:installation}

To install \texttt{SWIM}, the user needs to clone the GitHub repository to a convenient location on their system. The code has been developed and tested on Linux-based systems only, although it may be compatible with other platforms as well. We recommend creating a dedicated \texttt{conda} environment to ensure that all required dependencies are properly managed. After installing \texttt{Anaconda} (or \texttt{Miniconda}), the user needs to create a new environment using
\begin{verbatim}
conda create -n SWIM
\end{verbatim}
and activate it via
\begin{verbatim}
conda activate SWIM
\end{verbatim}
Then, they need to install the required packages within this environment:
\texttt{gcc}, \texttt{g++}, \texttt{gfortran}, \texttt{numpy}, \texttt{matplotlib}, \texttt{scipy}, \texttt{scikit-learn}, \texttt{jupyterlab}, \texttt{cffi}, and \texttt{joblib}. Additionally, it is required to install the \texttt{Boost} C++ libraries \footnote{\url{https://www.boost.org/releases/latest/}}, which are required for the numerical routines implemented in \texttt{SWIM}. For parallel execution, we recommend installing \texttt{OpenMPI} and enabling \texttt{OpenMP} support.

If parameter inference is required through MCMC analysis, \texttt{Cobaya} can be installed within the same environment by following the official installation instructions \footnote{\url{https://cobaya.readthedocs.io/en/latest/installation.html}}.

A shell script, \texttt{compile\_SWIM.sh}, is provided to compile the source code into the required shared libraries. The user must update the paths to the \texttt{SWIM} directory and the \texttt{Boost} installation within this script. The code should be recompiled whenever modifications are made to the \texttt{.cpp} source files. An overview of the submodules within \texttt{SWIM} is shown in Fig.~\ref{flowchart:SWIM_Overview}.

\section{Instructions to use $G(Q)$ submodule of \texttt{SWIM}} 
\label{appendix:GQ}

Before running the module, the user must implement the WI model in the files
\texttt{bg/model\_calc.cpp} and \texttt{pert/model\_calc.cpp}. This requires specifying:
\begin{itemize}
    \item In \texttt{bg/model\_calc.cpp}: $V(\phi)$ and $V_{,\phi}(\phi)$,
    \item In \texttt{pert/model\_calc.cpp}: $V(\phi)$, $V_{,\phi}(\phi)$, and $V_{,\phi\phi}(\phi)$,
\end{itemize}

For the dissipative coefficient:
\begin{itemize}
    \item If $\Upsilon \propto T^p \phi^c$, the user only needs to specify the exponents $(p,c)$ in the Python scripts, and all required derivatives are computed internally,

    \item Otherwise, the user must explicitly provide $\Upsilon(\phi,T)$ in both \texttt{bg} and \texttt{pert}, and its derivatives $\Upsilon_{,\phi}$ and $\Upsilon_{,T}$ in \texttt{pert/model\_calc.cpp}.
\end{itemize}

\begin{enumerate}
\item Input Specification for \texttt{find\_ICs.py}

The user must specify:
\begin{itemize}
    \item Model parameters (e.g., $V_0$, $g_*$), and dissipation parameters where applicable (e.g., $p$, $c$ for $\Upsilon \propto T^p \phi^c$),
    
    \item Inflation type via \texttt{hybrid\_inf}, and the critical field value if applicable,

    \item Target duration of inflation \texttt{dur\_N} (default: 60),

    \item Range and resolution of the dissipation ratio
    \begin{equation}
        Q \in [Q_{\min}, Q_{\max}],
    \end{equation}
    specified by \texttt{Qlow} (default: $10^{-9}$), \texttt{Qup} (default: $10^4$), and \texttt{npts} (default: 150),

    \item Bounds on the initial field value $\phi_i$ (in $\log_{10}(\phi/M_{\text{Pl}})$), for the root-finding algorithm,

    \item Number of processors \texttt{Nprocs} for parallel execution.
\end{itemize}

\item Input Specification for \texttt{find\_GQ.py}

The user must specify:
\begin{itemize}
    \item The same model and dissipation parameters as in \texttt{find\_ICs.py},

    \item Thermalization flag \texttt{therm}, where \texttt{therm}=1 enables thermal equilibrium (Bose--Einstein distribution) and \texttt{therm}=0 disables it,

    \item Radiation noise flag \texttt{rad\_noise}, where \texttt{rad\_noise}=1 includes the noise term in the radiation perturbation equation and \texttt{rad\_noise}=0 excludes it,

    \item Number of stochastic realizations \texttt{Nrealz} (default: 2048),

    \item Evaluation e-fold \texttt{Nstar} (default: 7), corresponding to the e-fold at which both numerical and analytical power spectra are evaluated.
\end{itemize}

\item Post-Processing

After generating \texttt{GQ.dat}, the user must:
\begin{itemize}
    \item Process and smooth the output using \texttt{GQ\_Plotting\_NB.ipynb},

    \item Save the final result as \texttt{GQ\_smooth.dat} for subsequent use in the semi-analytical module.
\end{itemize}
\end{enumerate}

\section{Instructions to use the semi-analytical power spectrum submodule of \texttt{SWIM}}
\label{appendix:semi-analytical-ps}

To use this module, the user should navigate to the \texttt{SA\_PS\_Calculator} subdirectory and specify the WI model in the \texttt{model\_calc.cpp} file. This module requires the $G(Q)$ correction factor to be precomputed. By default, this module uses the $G(Q)$ correction factor generated by the $G(Q)$ Calculator within \texttt{SWIM}. Alternatively, the user may provide a $G(Q)$ file generated using \texttt{WI2easy} (the path must be specified in the \texttt{GQfname} variable in \texttt{model\_calc.cpp}), or use analytical fits available in the literature, which can be defined in the \texttt{GQ} function within the \texttt{model\_calc.cpp} file.

The module also requires that \texttt{Cobaya} and the relevant cosmological likelihoods are installed and functional within the same \texttt{conda} environment used for \texttt{SWIM}. Two approaches are available for parameter inference:

\begin{itemize}
    \item \textbf{Constraints on $A_s$, $n_s$, and $r$:} This approach ignores correlations between these quantities and is computationally faster, as it does not require evaluation of the CMB power spectrum via \texttt{CAMB}. To use this mode, the files \texttt{llihood.py} and \texttt{Input\_asns.yaml} should be configured, with \texttt{Input\_asns.yaml} used as the input file for \texttt{Cobaya}.
    
    \item \textbf{Full CMB constraints:} This approach incorporates likelihoods such as Planck, ACT, SPT, and DESI. Although computationally more expensive, it provides more robust parameter constraints by computing the full CMB power spectrum from the primordial spectrum using \texttt{CAMB}. To use this mode, the files \texttt{An\_CAMB.py} and \texttt{Input.yaml} should be configured, with \texttt{Input.yaml} used as the input file for \texttt{Cobaya}.
\end{itemize}

The user must specify the following:

\begin{itemize}
    \item pivot scale exit can be computed automatically by setting \texttt{Np\_autocalc} = 1. If set to 0, the pivot scale exit must be specified explicitly or treated as a free parameter in the MCMC analysis.
    
    \item Source of $G(Q)$: If $G(Q)$ is provided as a precomputed data file (from either \texttt{SWIM} or \texttt{WI2easy}), set \texttt{read\_GQ\_from\_file} = 1. The code will internally interpolate this data to evaluate the semi-analytical power spectrum. If instead analytical fits are to be used, set \texttt{read\_GQ\_from\_file} = 0. Several commonly used fits are already implemented in the \texttt{GQ} function in \texttt{model\_calc.cpp}, which can also be extended to include user-defined forms (e.g., from \texttt{WarmSPy}).
    
    \item Choice of inference mode: For constraints using only $A_s$, $n_s$, and $r$, set \texttt{want\_full\_spectrum} = 0 in \texttt{llihood.py}. For full CMB-based constraints, set \texttt{want\_full\_spectrum} = 1 in \texttt{An\_CAMB.py}.
    
    \item Example \texttt{Cobaya} input files are provided for both inference modes.
\end{itemize}

\section{Instructions to use the numerical power spectrum submodule of \texttt{SWIM}} 
\label{appendix:numerical-ps}

To use this module, the user must navigate to the \texttt{PS\_Calculator} subdirectory and specify the WI model in the \texttt{model\_calc.cpp} file. For direct evaluation of the numerical power spectrum, the relevant scripts are located in the \texttt{Power\_Spectrum} directory, which contains the Python script \texttt{ps\_script.py}. This script computes the power spectrum as a function of $k$ for a given set of model parameters.

The user must specify:
\begin{itemize}
    \item the pivot scale $k_P$ in $\mathrm{Mpc}^{-1}$ (\texttt{kp}, default: $0.05$),
    \item the minimum step size for the stochastic differential equation solver (\texttt{em\_step}, default: $10^{-5}$),
    \item the number of realizations used to average the numerical power spectrum (\texttt{Nrealz}, default: 2048),
    \item the range of wavenumbers specified by \texttt{kmin} and \texttt{kmax}, which correspond to $\log_{10}(k_{\min})$ and $\log_{10}(k_{\max})$ (defaults: $-6$ and $2$, corresponding to $k \in [10^{-6}, 10^{2}]~\mathrm{Mpc}^{-1}$),
    \item the number of sampling points in this range (\texttt{points\_k}, default: 50),
    \item the e-fold corresponding to the pivot scale exit, $N_P$, which can either be computed internally by setting \texttt{Np\_autocalc}=1 using
    \begin{equation}
        k_P = e^{N_P - N_\mathrm{reh}} \frac{T_0}{T_\mathrm{reh}}
        \left( \frac{43}{11 g_s(T_\mathrm{reh})} \right)^{1/3} H(N_P),
    \end{equation}
    or specified explicitly by setting \texttt{Np\_autocalc}=0,
    \item whether to save the background evolution by setting \texttt{write\_bg = True}, in which case the evolution of $\phi(N)$, $\phi'(N)$, and $T(N)$ is written to \texttt{fname\_bg} (default: \texttt{bg.dat}),
    \item the output file name for the power spectrum (\texttt{fname\_ps}, default: \texttt{ps.dat}),
    \item the model parameters (e.g., $V_0$, $g_*$), along with flags controlling thermalization (\texttt{therm}) and inclusion of radiation noise (\texttt{rad\_noise}).
\end{itemize}

The script computes the numerical power spectrum by averaging over \texttt{Nrealz} stochastic realizations and outputs a table of $(k, P_{\mathcal{R}}(k))$. The accompanying notebook \texttt{Plotting\_NB.ipynb} can be used to visualize the spectrum and to fit it to a physically motivated parametrization of the form given in Eq.~(\ref{eq:fitting-ps}).

\subsection{Parameter Inference}

Parameter inference using the full numerical power spectrum, with Random Forest Regression (RFR) based emulation, is implemented in the \texttt{Emulator} module. This module consists of two components:

\begin{itemize}
    \item \texttt{RF\_Acc\_Cobaya}: Performs parameter inference using the full numerical power spectrum while training (and using) the RFR emulator on-the-fly within \texttt{Cobaya}.
    
    \item \texttt{RF\_Only\_Cobaya}: Performs parameter inference using a pre-trained RFR emulator. In this mode, the full numerical power spectrum is not recomputed, resulting in significantly faster evaluations.
\end{itemize}

The WI model is specified in \texttt{model\_calc.cpp} within the \texttt{PS\_Calculator} module and is shared by the emulator-based inference routines. To perform inference, the user must configure the provided \texttt{Cobaya} input file (\texttt{Input\_asns.yaml}) along with the likelihood implementation in \texttt{llihood\_Observables.py}.

\acknowledgments

The work of S.D. is supported by the Start-up Research Grant (SRG) awarded by Anusandhan National Research Foundation (ANRF), Department of Science and Technology, Government of India 
[File No. SRG/2023/000101/PMS]. UK warmly acknowledges the Axis Bank PhD program at Ashoka University for the PhD fellowship provided by Axis Bank. UK is grateful to Gabriele Montefalcone, Alejandro Perez Rodriguez, and Dipankar Bhattacharya for insightful discussions that have significantly contributed to this work.

After the first version of this manuscript appeared on arXiv, we received correspondences from the authors of \texttt{WI2easy} that help up understand the reason behind the discrepancies observed in \texttt{SWIM}'s $G(Q)$ and \texttt{WI2easy}'s $G(Q)$ in the moderate dissipative regime. We thus thank the authors of \texttt{WI2easy}, especially Rudnie O. Ramos, for their insightful discussions. The authors would also like to thank the anonymous referee for their insightful comments which we believe have helped enhance the presentation of the results of the manuscript. 




\bibliographystyle{JHEP}
\bibliography{refs}

\end{document}